\documentclass[journal,transmag]{IEEEtran}
\ifCLASSINFOpdf
\else
\fi

\usepackage{graphicx}
\usepackage{epsfig}
\usepackage{epstopdf}
\usepackage{amsthm,amsfonts,amssymb}
\usepackage{amsmath}
\usepackage{amssymb}
\usepackage{subfig}
\usepackage{tabularx}
\usepackage{cases}
\usepackage{mathrsfs}
\usepackage{etoolbox}
\usepackage{url}
\usepackage{bm}
\usepackage{multirow}
\usepackage{booktabs}
\usepackage[numbers]{natbib}
\usepackage{enumerate}
\usepackage{float}
\usepackage[usenames, dvipsnames]{color}

\theoremstyle{plain}
\newtheorem{thm}{Theorem}
\theoremstyle{plain}

\theoremstyle{plain}

\theoremstyle{plain}

\theoremstyle{plain}

\theoremstyle{plain}

\theoremstyle{definition}

\theoremstyle{definition}

\theoremstyle{definition}

\theoremstyle{definition}

\theoremstyle{remark}

\theoremstyle{remark}

\begin{document}
%
\title{Multi-agent Inverse Reinforcement Learning for Two-person Zero-sum Games}


\author{\IEEEauthorblockN{Xiaomin Lin\IEEEauthorrefmark{1},
Peter A. Beling\IEEEauthorrefmark{1}, \emph{Member, IEEE}, and
Randy Cogill\IEEEauthorrefmark{2}}
\IEEEauthorblockA{\IEEEauthorrefmark{1}Department of Systems and Information Engineering,
University of Virginia, Charlottesville, VA 22903 USA}
\IEEEauthorblockA{\IEEEauthorrefmark{2}IBM Research Dublin, Dublin, Ireland}
\thanks{Manuscript received XX; revised XX. 
Corresponding author: Peter A. Beling (email: pb3a@virginia.edu.)}}

\markboth{IEEE Transactions on Computational Intelligence and AI in Games,~Vol.~XX, No.~XX, December~XX}%
{Shell \MakeLowercase{\textit{et al.}}: Bare Demo of IEEEtran.cls for Journals}
%



\IEEEtitleabstractindextext{%
\begin{abstract}
The focus of this paper is a Bayesian framework for solving a class of problems termed {\em multi-agent inverse reinforcement learning} (MIRL). Compared to the well-known {\em inverse reinforcement learning} (IRL) problem, MIRL is formalized in the context of stochastic games, which generalize Markov decision processes to game theoretic scenarios. We establish a theoretical foundation for competitive two-agent zero-sum MIRL problems and propose a Bayesian solution approach in which the generative model is based on an assumption that the two agents follow a minimax bi-policy. Numerical results are presented comparing the Bayesian MIRL method with two existing methods in the context of an abstract soccer game. Investigation centers on relationships between the extent of prior information and the quality of learned rewards. Results suggest that covariance structure is more important than mean value in reward priors.
\end{abstract}

}

\maketitle

\IEEEdisplaynontitleabstractindextext

%
\IEEEpeerreviewmaketitle

\section{Introduction}\label{section_1}
\IEEEPARstart{L}{earning} from demonstrations (LD) is a traditional line of research in behavior learning, and is particularly useful in game design. In LD, policy learning directly from observations has achieved remarkable success in large part because it can benefit from advanced  supervised learning techniques. Examples of this point can be seen in \cite{Runarsson2014}, where preference learning is used for policy learning, and in \cite{Maddison2015}, where a deep convolutional neural network is adopted as the basis for policy learning. In recent years, reinforcement learning (RL) has attracted the interest of game designers  because it aligns with the belief that behavior is mainly reward-driven (see, e.g., \cite{Wang2010,Glavin2015}). Inverse reinforcement learning (IRL), which aims to recover reward (equivalently, payoff or cost) functions given measurements of an agent’s behavior over time as well as a model of the environment, was introduced  in \cite{Russell1998} and then formalized in \cite{Ng2000} in the context of several linear programming algorithms. One can view the IRL problem as being that of learning the reward structure for a game given observations of the play of an expert. One major advantage of IRL, as pointed out in \cite{Ng2000}, is that in many applications, the reward function provides a parsimonious description of behavior that is succinct, robust, and transferable with respect to changes in the environment. In comparison to policy learning, the benefit of IRL to the game community is that it has the potential to yield insights into the value systems driving player behavior, which in turn might help designers balance the difficulty of the game play and tune the user experience.    
\par
One shortcoming of IRL that is particularly relevant to games is that it assumes no other adaptive agents exist in the environment. However, many games are multi-agent, mutually influential systems. To jointly consider the decision making processes of interacting rational agents requires different models and techniques. In the forward direction, \emph{multi-agent reinforcement learning} (MRL), proposed by Littman \cite{Littman1994}, extends RL to a multi-agent framework. Littman makes use of stochastic games (see, e.g., \cite{Owen1968}) to model MRL, limiting consideration to the special case of \emph{two-player zero-sum games}, in which one agent's gain is always the other's loss, and applies this algorithm in a simple grid-world soccer game. Hu and Wellman \cite{Hu1998} extend Littman's work, proposing a \emph{two-player general-sum} stochastic game framework for the MRL problem. They point out that the concept of optimality loses its meaning in MRL problems since any agent's payoff depends on the action choices of others. Consequently, they adopt as a solution concept the \emph{Nash equilibrium}, in which each agent's choice is the best response to other agents' choices. Later MRL work has focused on the development of solution concepts and methods, in competing games as well as cooperative games, including \cite{Abdallah2008,Ghavamzadeh2006,Patek2007,Zhao2008}. Representative applications include   traffic control \cite{Bazzan2009} and robotics  \cite{Duan2012}. 
\par
The inverse learning problem for MRL, or {\em multi-agent inverse reinforcement learning} (MIRL), is more complicated than IRL. In the context of a stochastic game being played by two or more agents, the problem is that of estimating the game payoffs given observations of the actions taken by the players and the state transitions. Games bring two primary challenges relative to the Markov decision processes (MDPs) used in IRL. First, as Hu and Wellman \cite{Hu1998} note, the concept of optimality, central to MDPs, must be replaced with an equilibrium solution concept, such as the Nash equilibrium. Second, the non-uniqueness of equilibrium strategies (especially for two-player non-zero sum games) means that in MIRL, in addition to multiple reasonable solutions for a given inversion model, theoretically there may be multiple inversion models that are all equally sensible approaches to solving the problem. 
\par
This paper proposes a novel Bayesian approach to MIRL. We establish a theoretical foundation for competitive two-agent zero-sum MIRL problems and describe {\em Bayesian MIRL} (BMIRL), a Bayesian solution approach  in which the generative model is based on an assumption that the two agents follow a minimax bi-policy. To our knowledge, this topic has not been deeply studied in the literature. Natarajan {\em et al.} \cite{Natarajan2010} present an inverse reinforcement learning model for multiple agents. However, that paper does not consider competing agents or game-theoretic models, a key characteristic of our work. Waugh \emph{et al.} \cite{Waugh2011} do consider a form of the inverse equilibrium problem. However, that paper considers simultaneous one-stage games, rather than the sequential stochastic games we consider here. A similar method, termed {\em decentralized MIRL} (d-MIRL), is a decentralized linear IRL approach based off work by Reddy {\em et al.} \cite{Reddy2012}, while our work is centralized and set in a Bayesian framework. 
\par
Several numerical experiments are performed in the setting of an abstract soccer game with simple grid structure and movement actions and probability models governing ball exchange and the outcomes of ball kicks at the goal (the agents' {\em shoot} action). For the inverse learning problem, the unknown rewards correspond to location of goals and player perception of a successful shot from each position on the field.  Investigation centers on relationships between the extent of prior information and the quality of learned rewards. The quality of learned rewards is measured by distance metrics in reward and probability space and by the game playing success of agents that use the rewards as the basis for an equilibrium policy. The weakest priors result in learned rewards that would give an agent using them no chance of winning the game, while the strongest priors result in learned rewards essentially as good as ground truth. Additionally, results suggest that covariance structure is more important than mean value in reward priors.
\par
The remainder of the paper is structured as follows: Section \ref{section_2} introduces notation, terminology, definitions, and some basic properties needed for later work. Section \ref{section_3} provides the main technical results, including a Bayesian framework for MIRL and formulation of a convex optimization problem for learning rewards. Section \ref{section_4} and \ref{section_5} extend the BIRL and d-MIRL approaches to the case where reward is also action dependent. Section \ref{section_6} introduces the soccer model and compares the results generated from the three methods. Section \ref{section_7} provides evaluation of learned rewards of our BMIRL method in terms of game playing success in simulations of the soccer game. Section \ref{section_8} and \ref{section_9} offer concluding remarks and a discussion of future work, respectively.
\par
\section{Preliminaries}\label{section_2}
\subsection{Stochastic Games}
A two-player \emph{discounted stochastic game} is played as follows. The game begins in one of finitely many states. There is a reward for each player. In each state, each player simultaneously selects one of finitely many actions, and hence receives a reward that associates with current state and sometimes, as well as the actions selected by one or both players. The game then makes a stochastic transition to a new state, where the transition is dependent on the starting state and the jointly selected actions. This process is repeated over an infinite time horizon, where geometrically discounted rewards are accrued additively.
\par
Under these rules, we can specify an instance of a two-person zero-sum discounted stochastic game in terms of the state space $\mathcal{S}=\left \{ 1,2,\cdots ,N \right \}$, the action spaces $\mathcal{A}_1=\mathcal{A}_2=\left \{ 1,2,\cdots ,M \right \}$, two reward vectors ${r}^1$ and ${r}^2$ of the two agents involved, state transition probabilities $p\left ( s'|s, a^1, a^2 \right )$, and a reward discount factor $\gamma \in \left [ 0, 1 \right )$. Reward values are assumed to be dependent on state and the actions taken by the two agents. Hence, the dimension of $r^1\left ( r^2 \right )$ depends on the the size of $S$, $\mathcal{A}_1$ and $\mathcal{A}_2$. We will use $r^1\left ( \cdot  \right )\left ( r^2\left ( \cdot  \right ) \right )$ to denote a scalar; e.g., $r^1\left ( s, a^1, a^2  \right )$ represents the reward value gained by agent 1 when the two agents take actions  $a^1$ and $a^2$, respectively,  in state $s$. 
\par
A solution to a stochastic game is a \emph{bi-policy}, which provides the rules that each player follows when selecting actions at each state. Without loss of generality, a bi-policy can be specified by a collection of conditional probability mass functions ${\pi}^1$ and ${\pi}^2$, where player $k$ selects action $a^k$ in state $s$ with probability $\pi^k(a^k | s)$. Each $\pi^k( \cdot | s)$ is referred to as the \emph{strategy} played by player $k$ in state $s$.
\par
Given that each player can select from among $M$ actions, the strategy followed by player $k$ in state $s$ can be represented by the $M \times 1$ vector $\pi^k\left ( s \right )$. The si for state $s$ is the set of two column vectors that denote the strategies employed by player 1 and player 2 in state $s$,
\begin{equation*}
\pi\left ( s \right )=\left \{ \pi^1\left ( s \right ), \pi^2\left ( s \right ) \right \}.
\end{equation*}
In this notation, the bi-policy is defined as the set of all bi-strategies over all states,
\begin{equation*}
\pi=\left \{ \pi\left ( 1 \right ),\pi\left ( 2 \right ),\cdots ,\pi\left ( N \right ) \right \}.
\end{equation*}
\subsection{Zero-sum Case} \label{sec:zerosum}
A \emph{two-player zero-sum discounted stochastic game} is a special case of the game defined above for which $r^1\left ( s, a^1, a^2 \right ) = -r^2\left ( s, a^1, a^2 \right )$. The symmetry of rewards between the two players allows to use $r$ to denote $r^1$.  Attention is restricted to the zero-sum case throughout the remainder of the paper.
\par
We use $\tilde{r}_{\pi}\left ( s \right )$ to denote the single-stage \emph{expected reward value} received by agent $1$ at state $s$ under bi-policy $\pi$. Then $\tilde{r}_{\pi}$ is a column vector with its $i$th component $\tilde{r}_{\pi}\left ( s \right )$. Define $\tilde{r}_{\pi}\left ( s \right )$ to be
\begin{equation}\label{R_average}
\begin{aligned}
\tilde{r}_{\pi}\left ( s \right ) &=  \sum_{a^1, a^2}\pi^1\left ( a^1 |s\right )\pi^2\left ( a^2|s \right )r\left ( s, a^1, a^2 \right ) \\
&=  \left [ \pi^1\left ( s \right ) \right ]^Tr\left ( s \right )\pi^2\left ( s \right ),
\end{aligned}
\end{equation}
where $r\left ( s \right )$ is a $M \times M$ matrix, whose entries are independent of $\pi\left ( s \right )$.
We can express this relationship in matrix notation as
\begin{equation}\label{r_ave}
\tilde{r}_{\pi}=B_{\pi}r,
\end{equation}
where $B_{\pi}$ is a $N \times NM^2$ matrix constructed from bi-policy $\pi$, whose $k$th row is:
\begin{equation*}
\left [ \Phi^{\pi}_{1,1}\left ( k \right ), \Phi^{\pi}_{1,2}\left ( k \right ),\cdots ,\Phi^{\pi}_{M,M}\left ( k \right )   \right ],
\end{equation*}
where 
\begin{equation*}
\Phi^{\pi}_{i,j}\left ( k \right ) = \left [ \underbrace{0,\cdots ,0}_{k-1},\phi^{\pi}_{i,j}\left ( k \right ), \underbrace{0,\cdots ,0}_{N-k} \right ],
\end{equation*}
and 
\begin{equation*}
\phi^{\pi}_{i,j}\left ( k \right )=\pi^1\left ( i|k \right )\pi^2\left ( j|k \right ).
\end{equation*}
\par
The concepts of the \emph{value function} and $Q$-\emph{function} in MDPs have natural analogs in zero sum stochastic games. In particular, let us define the value function to be the bi-policy-dependent, discounted expected sum of rewards of player $1$ as a function of the initial state $s$:
\begin{equation}\label{V_orginial}
V_{\pi}\left ( s \right ) = \sum_{t=0}^{\infty }\gamma^tE\left ( \tilde{r}_{\pi}\left ( s_t \right ) |s_0=s \right ),
\end{equation}
where $s_t$ denotes the state of the game at stage $t$ and $\tilde{r}^t_{\pi}$ denotes player $1$'s expected reward under bi-policy $\pi$ at that stage. Note that the superscript $t$ can be removed because of the Markov property. $V_{\pi}$ denotes the column vector with $i$th component $V_{\pi}\left ( i \right )$. 
\par
In addition, we define player $1$'s Q-function of state $s$ and action pair $\left ( a^1, a^2 \right )$, under bi-policy $\pi$, as
\begin{equation}\label{Q_element}
Q_{\pi}\left ( s,a^1, a^2 \right )=r\left ( s, a^1, a^2 \right ) + \gamma \sum_{s'}p\left ( s'|s, a^1, a^2 \right )V_{\pi}\left ( s' \right ).
\end{equation}
Over all states and actions, we can write equation \eqref{Q_element} in matrix notation as
\begin{equation}\label{Q_all}
Q_{\pi}=r+\gamma P V_{\pi}, 
\end{equation}
where $P$ is a $NM^2 \times N$ matrix with $p\left ( s'|s, a^1, a^2 \right )$ as its elements.
\par
Let $G_{\pi}$ denote transition matrix under bi-policy $\pi$. Specifically, $G_{\pi}$ is the $N \times N$ matrix with elements
\begin{equation}\label{G_def}
g_{\pi}\left ( s' | s \right )=\sum_{a^1, a^2}\pi^1\left ( a^1|s \right )\pi^2\left ( a^2|s \right )p\left ( s'|s, a^1, a^2 \right ).
\end{equation}
\par
Note that 
\begin{equation}\label{V_expend}
\begin{aligned}
V_{\pi}\left ( s \right ) &= \tilde{r}_{\pi}\left ( s \right ) +  \sum_{t=1}^{\infty }\gamma^{t}E\left ( \tilde{r}_{\pi}\left ( s_t \right ) |s_0=s \right )  \\
&=\tilde{r}_{\pi}\left ( s \right ) + \gamma\sum_{s'}g_{\pi}\left ( s'|s \right )V_{\pi}\left ( s' \right ).
\end{aligned}
\end{equation}
This equation can be written in matrix notation as
\begin{equation}\label{V_comp}
V_{\pi}=\tilde{r}_{\pi}+\gamma G_{\pi}V_{\pi}.
\end{equation}
Thus
\begin{equation}\label{V_comp2}
V_{\pi}=\left ( I - \gamma G_{\pi} \right )^{-1}B_{\pi}r,
\end{equation}
where $\left ( I-\gamma G_{\pi} \right )$ is always invertible for $\gamma \in \left [0, 1  \right )$ since $G_{\pi}$ is a transition matrix.
The value function $V_{\pi}\left ( s \right )$ can be expressed in terms of the $Q$-function as
\begin{equation}\label{shapley_formula}
V_{\pi}\left ( s \right ) = \left [ \pi^1\left ( s \right ) \right ]^T Q_{\pi}\left ( s \right )\pi^2\left ( s \right ),
\end{equation}
where $Q_{\pi}\left ( s \right )$ is a $M \times M$ matrix for agent $1$, whose $\left ( i, j \right )$ element is given by $Q_{\pi}\left ( s, i, j \right )$. Note that while $Q_{\pi}\left ( s \right )$ is a matrix, the $Q_{\pi}$ introduced in \eqref{Q_all} is an $NM^2 \times 1$ vector. We will use this relationship between the Q-function and the value function to define a \emph{minimax bi-policy} for a stochastic game.
\par
We will assume that rational agents playing two-player zero-sum stochastic games seek a minimax bi-policy. A minimax bi-policy is an equilibrium, in that it has the property that neither player can change the game value in their favor given that the other player holds their policy fixed. To give a precise definition of a minimax bi-policy, we will start by reviewing the notion of a minimax bi-strategy for a static game \cite{Neumann1944}.
\par
First consider a static (single-stage) zero-sum game, where two players simultaneously choose an action and both players receive a reward determined by the joint choice of actions. The minimax theorem states that for every two-person zero-sum game with finitely many actions, there exists a value $V$ and a mixed strategy for each player such that
\begin{itemize}
\item Given player 2\textquoteright s strategy, the best expected reward possible for player 1 is $V$.
\item Given player 1\textquoteright s strategy, the best expected reward possible for player 2 is $-V$.
\end{itemize}
\par
As before, the strategies played by both players in a certain state $s$ can be expressed in terms of probability mass functions $\pi^1\left ( s \right )$ and $\pi^2\left ( s \right )$. Expressing the reward received by player $1$ as an $M \times M$ matrix $Q_{\pi}\left ( s \right )$, the value of the game for player $1$ under a minimax bi-strategy is given by
\begin{equation*}
\mbox{value}\left ( Q_{\pi}\left ( s \right ) \right )=\underset{\pi^1\left ( s \right )}{\mbox{max}}\left \{ \underset{\pi^2\left ( s \right )}{\mbox{min}}\left \{ \left [ \pi^1\left ( s \right ) \right ]^T Q_{\pi}\left ( s \right ) \pi^2\left ( s \right ) \right \} \right \}.
\end{equation*}
\par
A pair $\pi^1\left ( s \right )$ and $\pi^2\left ( s \right )$ that achieves this value is called a \emph{minimax bi-strategy}. For zero-sum games, a minimax bi-strategy is also a Nash equilibrium.
\par
The concept of a minimax bi-strategy can be extended to two-player discounted stochastic games via the following theorem \cite{Shapley1953}.
\begin{thm}[Shapley's Theorem]\label{sharpley}
There exists a bi-policy $\pi$ such that
\begin{equation}
V_{\pi}\left ( s \right ) = \emph{\mbox{value}}\left ( Q_{\pi}\left ( s \right ) \right )
\end{equation}
for all $s \in \mathcal{S}$.
\end{thm}
A bi-policy that satisfies Theorem \ref{sharpley} is called a \emph{minimax bi-policy}. For a minimax bi-policy, $V_{\pi}\left ( s \right )$ gives the game value from each initial state $s \in \mathcal{S}$. Throughout the following sections it is assumed that agents are observed playing a game according to a minimax bi-policy and that the complete bi-policy is observable.  The minimax nature of the bi-policy can then be used to infer the reward structure of the game. 
\par
\section{Bayesian MIRL}\label{section_3}
We will formulate two-agent MIRL problems in a Bayesian  setting. Bayesian methods have been widely adopted for IRL problems \cite{Baker2009,Choi2011,Dimitrakakis2011,Engel2005,Michini2012,Qiao2011,Ramachandran2007}. In a Bayesian setting, we assign a prior distribution to the reward functions. This prior distribution encodes the learner's initial belief about the reward functions before any observations are made. 

 Given an observed bi-policy, we can generate a point estimate of the reward function from the posterior distribution over reward functions. To construct this point estimate, we must know the likelihood of observing each bi-policy for each given reward function. So, consideration must be given to determining the appropriate likelihood function for the MIRL problem and to the development of optimization models that can be used to generate point estimates of the reward function.
\par
The BMIRL approach we propose is a maximum a posteriori probability (MAP) estimate of reward under a likelihood function that encodes the notion of a minimax equilibrium. Let $f\left ( r \right )$ denote the prior distribution on the reward of agent $1$ (recalling that we denote $r = r^1$ and $r^1 = -r^2$ for zero-sum games). We will discuss the selection of prior distributions further in Section 3.1. Also, let $p\left ( \pi|r \right )$ denote the likelihood of observing bi-policy $\pi$ when the true reward is $r$. Hence now our objective is to maximize $f\left ( r|\pi \right )$, the posterior of rewards given an observing bi-policy, as follows,
\begin{equation*}
f\left ( r| \pi \right ) \propto p\left ( \pi|r \right )f\left ( r \right ).
\end{equation*}
\par
\subsection{Prior Distributions on Rewards}\label{subsec:prior}
In BMIRL, we use prior distributions over reward functions to model our initial uncertainty in the reward. Although any prior may be used, in this paper we prefer Gaussian priors for rewards. Gaussians are a reasonable choice of prior since they provide a straightforward model for representing uncertainty around a nominal choice of reward function, and have the added benefit of leading to analytically tractable inference procedures.
\par
Specifically, we model ${r}\sim \mathcal{N}\left ( {\mu_r}, \Sigma_{{r}} \right )$, where ${\mu_r}$ is the mean of ${r}$ and $\Sigma_{{r}}$ is the covariance matrix. The probability density function of ${r}$ is
\begin{equation}\label{r_density}
f\left ( {r} \right )= \frac{1}{\left ( 2\pi \right )^{N/2}\left | \Sigma_{{r}} \right |^{1/2}}\exp\left ( -\frac{1}{2}\left ( {r-{\mu_r}} \right )^T \Sigma^{-1}_{{r}} \left ( {r-{\mu_r}} \right ) \right ).
\end{equation}
\subsection{Likelihood Function (Unique Minimax bi-policy)}
To model the likelihood function $p(\pi|r)$, we assume that the bi-policy which the two agents follow is a unique minimax bi-policy given $r$. The likelihood is then a probability mass function given by
\begin{equation}
p\left ( \pi| r \right )=\begin{cases}
1, &  \mbox{if }\pi\mbox{ is minimax for }r \\
0, & \mbox{otherwise.}
\end{cases}
\end{equation}
\par
\subsection{MAP Estimation Model}
The posterior distribution of rewards for a given observed
bi-policy is now
\begin{equation*}
f\left ( r| \pi \right ) \propto p\left ( \pi|r \right )f\left ( r \right ) = \begin{cases}
f\left ( r \right ), &  \mbox{if }\pi\mbox{ is minimax for }r \\
0, & \mbox{otherwise}.
\end{cases}	
\end{equation*}
The MAP estimate of rewards is the vector $r$ that maximizes $f\left ( {r}|{\pi} \right )$. Thus we wish to solve the problem
\begin{equation}\label{model}
\begin{aligned}
\text{maximize:} \quad
&f\left ( {r} \right ) \\
\text{subject to:} \quad
&p\left ( {\pi}| {r} \right ) = 1.
\end{aligned}
\end{equation}
The remainder of this section will be devoted to developing a tractable characterization of the set of feasible ${r}$.   Consider, as a first step, the class of static, single-stage, zero-sum games. In these games, minimax strategies satisfy the conditions of the following theorem \cite{Neumann1944,Ferguson2008}.
\begin{thm}[Minimax Theorem]\label{thm4}
Consider a two-person zero-sum game with $M\times M$ payoff matrix $A$. There exists a value $V$, a mixed strategy $p$ for player 1, and a mixed strategy $q$ for player 2 such that 
\begin{equation}\label{corollary_minimax}
\begin{aligned}
 A^{T}p & \geq V{1}_{M}\\
 Aq & \leq V{1}_{M},
\end{aligned}
\end{equation}
where ${1}_{M}$ is the $M \times 1$ vector in which every element is $1$. Moreover, $p$ and $q$ are an equilibrium bi-strategy and $V$ is the game value if and only if \eqref{corollary_minimax} holds.
\end{thm}
\par
This theorem has direct implications for inverse learning problems. Consider a static game as a special case of the MIRL problem, where the goal is to recover a $A$ such that the given bi-strategy $\left ( p, q \right )$ is a minimax bi-strategy. Hence, the linear constraints \eqref{corollary_minimax} give a characterization of the desired constraint set for a two-person zero-sum static game.
\par
We will now extend this approach to a multi-stage stochastic game. Combining Theorem \ref{sharpley} with Theorem \ref{thm4}, a bi-policy ${\pi}$ is a minimax bi-policy if and only if
\begin{equation}\label{corollary_bi-policy}
\begin{aligned}
\left [ Q_{{\pi}}\left ( s \right ) \right ]^T\pi^1\left ( s \right ) &\geq V_{{\pi}}\left ( s \right ){1}_M\\
Q_{{\pi}}\left ( s \right )\pi^2\left ( s \right )&\leq V_{{\pi}}\left ( s \right ){1}_M,
\end{aligned}
\end{equation}
for all $s \in \mathcal{S}$. The linear inequalities \eqref{corollary_bi-policy} provide conditions that must hold for the $Q$-function and value function of a stochastic game if ${\pi}$ is a minimax bi-policy.
\par
Since our ultimate goal is to estimate the reward function of a stochastic game, we must introduce additional constraints relating the $Q$-function and value function to rewards. From \eqref{Q_all} and \eqref{V_comp2}, recall that
\begin{equation}\label{Q_and_V}
\begin{aligned}
& Q_{{\pi}}={r}+\gamma P V_{{\pi}} \\
& V_{{\pi}}=\left ( I - \gamma G_{{\pi}} \right )^{-1}B_{{\pi}}{r},
\end{aligned}
\end{equation}
and from \eqref{R_average}, \eqref{r_ave} and \eqref{shapley_formula}, we can deduce that
\begin{equation}\label{V_Q_B}
V_{{\pi}}=B_{{\pi}}Q_{{\pi}}.
\end{equation}
Let $B_{{\pi}^1|a^2=j}$ denote the $B_{{\pi}}$ obtained when ${\pi}^1$ is used as player 1's policy, and player 2 selects action $a^2 = j$ in all states. In this notation, the inequalities \eqref{corollary_bi-policy} can be expressed as
\begin{equation}\label{B_Q_inequalities}
\begin{aligned}
& B_{{\pi}^1|a^2=j}Q_{{\pi}}\geq B_{{\pi}}Q_{{\pi}}, \forall j \in \mathcal{A}_2 \\
& B_{{\pi}^2|a^1=i}Q_{{\pi}}\leq B_{{\pi}}Q_{{\pi}}, \forall i \in \mathcal{A}_1. \\
\end{aligned}
\end{equation}
Substituting the expression for $V_{{\pi}}$ into the expression for $Q_{{\pi}}$ in \eqref{Q_and_V}, we obtain
\begin{eqnarray}\label{Q_complicated}
Q_{{\pi}} &=& {r} + \gamma P\left ( I - \gamma G_{{\pi}} \right )^{-1}B_{{\pi}}{r}\\
&= &\left ( I + \gamma P\left ( I - \gamma G_{{\pi}} \right )^{-1}B_{{\pi}} \right ){r}.
\end{eqnarray}
Finally, letting
\begin{equation}\label{D_def}
D_{{\pi}} = \left ( I + \gamma P\left ( I - \gamma G_{{\pi}} \right )^{-1}B_{{\pi}} \right ), 
\end{equation}
the inequalities \eqref{B_Q_inequalities} can be expressed as
\begin{equation}\label{proposition_constraints_inequalities}
\begin{aligned}
&\left ( B_{{\pi}^1|a^2=j}-B_{{\pi}} \right )D_{{\pi}}{r}\geq 0,\forall j\in\mathcal{A}_2 \\
&\left ( B_{{\pi}^2|a^1=i}-B_{{\pi}} \right )D_{{\pi}}{r}\leq 0,\forall i\in\mathcal{A}_1. \\
\end{aligned}
\end{equation}
\par
Now we can formulate a convex quadratic program equivalent to \eqref{model}. Recall that we use a Gaussian prior in this paper, so the objective function in \eqref{model} is log-concave. To obtain an equivalent convex optimization problem, we will instead minimize $-\ln \left ( f\left ( {r} \right ) \right )$.
Combining \eqref{proposition_constraints_inequalities} with the negative log-prior objective, the optimization problem \eqref{model} can
be solved as the following equivalent convex quadratic program:
\begin{equation}\label{complex_program}
\begin{aligned}
\textrm{minimize:}  \quad
& \frac{1}{2}\left ( {r-{\mu_r}} \right )^T {\Sigma_r}^{-1} \left ( {r-{\mu_r}} \right )   \\
\textrm{subject to:} \quad
&\left ( B_{{\pi}^2|a^1=i}-B_{{\pi}} \right )D_{{\pi}}{r}\leq {0}\\
\quad
&\left ( B_{{\pi}^1|a^2=j}-B_{{\pi}} \right )D_{{\pi}}{r}\geq {0},\\
\end{aligned}
\end{equation}
for all $i\in\mathcal{A}_1$ and $j\in\mathcal{A}_2$.
\par
The optimization problem \eqref{complex_program} is specific to two-person zero-sum MIRL problems, which is a class of problems in which the reward value depends on both state and bi-actions. The equivalent problem for the case where reward values only depend on state is as follows:
\begin{equation*}
\begin{aligned}
\textrm{minimize:}  \quad
& \frac{1}{2}\left ( {r}-\mu_{{r}} \right )^T \Sigma_{{r}}^{-1}\left ( {r}-\mu_{{r}} \right )  \\
\textrm{subject to:}  \quad
&\left ( G_{{\pi}}-G_{{\pi}^2|a^1=i} \right )\left ( {I}-\gamma G_{{\pi}} \right )^{-1}{r} \geq {0} \\
                          &\left ( G_{{\pi}}- G_{{\pi}^1|a^2=j} \right )\left ( {I}-\gamma G_{{\pi}} \right )^{-1}{r} \leq {0} \\
\end{aligned}
\end{equation*}
for all $i\in\mathcal{A}_1$ and $j\in\mathcal{A}_2$.
\par
It is worth discussing the scalability of the optimization problem. When the problem size, $n$, is large, the inversion of the covariance matrix, which is usually sparse, is computationally expensive ($O(n^3)$). And even if we obtain the inverse of the covariance matrix (which generally will not be sparse), the objective of this problem includes $O(n^2)$ quadratic monomials, which may not fit into memory. One way to tackle this problem is to first compute the Cholesky upper-triangle factorial $R$ of $\Sigma$, which often is sparse as $\Sigma$ itself is sparse. Then let $e = R^{T}\left ( r - \mu_r \right )$ and add it to the constraints. Finally, we rewrite our objective as $\frac{1}{2}e^{T}e$. This reformulation helps avoid the memory issue.

\subsection{Discussion on Nonunique bi-policies}
In the definition of the likelihood function and the convex program \eqref{complex_program} we have implicitly assumed that the stochastic game has a unique minimax bi-policy.  It is important to note that this assumption need not hold.  Indeed for a {\em static}  two-person zero-sum games there may exist an infinite number of minimax bi-strategies, even though each such game has a unique Nash equilibrium value. In \cite{Ferguson2008}, a sufficient condition for the existence of unique bi-strategy for a static matrix game is given: the square game matrix $A$ is nonsingular and ${1}^TA^{-1}{1}\neq 0$. 
 
So it is clear we must consider cases where multiple minimax bipolices exist. For ease of expression in doing so,  define the following notation:
\begin{itemize}
\item $\mathcal{G}\left ( r \right )$: the stochastic game given one agent's reward vector is $r$.
\item $\mathcal{U}\left ( r \right )$: the set of $r$ in which the necessary condition \eqref{B_Q_inequalities} is satisfied.
\item $\mathcal{U}^{*}\left ( r \right )$: a subset of $\mathcal{U}\left ( r \right )$ where $\pi$ is a unique minimax bi-policy for $\mathcal{G}\left ( r \right )$.
\item $\mathcal{M}\left ( \mathcal{U}\left ( r \right ) \right )$: the optimization problem \eqref{complex_program} where $r \in \mathcal{U}\left ( r \right )$.
\item $\mathcal{M}\left ( \mathcal{U}^{*}\left ( r \right ) \right )$: a subproblem of \eqref{complex_program} constrained by $r \in \mathcal{U}^{*}\left ( r \right )$.
\end{itemize}
We would like to solve the MAP problem for $\mathcal{G}\left ( r \right )$ that accounts for the possibility of multiple minimax strategies.  Even with a generative notion such as the idea that agents will select among equal-value equilibrium strategies with uniform probability, however, it is difficult to develop a likelihood for this problem because we cannot easily characterize the set of minimax equilibrium strategies as a function of $r$.  As a surrogate, one might adopt $\mathcal{M}\left ( \mathcal{U}^{*}\left ( r \right ) \right )$, but again this problem is difficult to define directly.  An alternate approach is to first solve $\mathcal{M}\left ( \mathcal{U}\left ( r \right ) \right )$.  Let $\tilde{r}$ be the optimal solution to this problem. If $\tilde{r} \in {U}^{*}$ then $\tilde{r}$ is optimal for  $\mathcal{M}\left ( \mathcal{U}^{*}\left ( r \right ) \right )$. If $\tilde{r} \not \in {U}^{*}$ then form $\hat{r} = \tilde{r}+\epsilon	 $, for small random perturbation $\epsilon$.  With high probability $\hat{r} \in {U}^{*}(r)$ (cf. \cite{Rudelson2014}) and will be nearly optimal for $\mathcal{M}\left ( \mathcal{U}^{*}\left ( r \right ) \right )$.
\subsection{Uniqueness of bi-policy}\label{sec:2_3}
For a {\em static}  two-person zero-sum games there may exist multiple minimax bi-strategies, even though each such game has a unique Nash equilibrium. In \cite{Ferguson2008}, a sufficient condition for the existence of unique bi-strategy for a matrix game is given: the square game matrix $A$ is nonsingular and ${1}^TA^{-1}{1}\neq 0$. Note that this is not the necessary condition for the existence of unique minimax bi-strategy. Rudelson and Vershynin \cite{Rudelson2014} show that a perturbation of any fixed square matrix by a random unitary matrix is well invertible with high probability. From these findings, we can come to conclusion that in a real world two-person zero-sum MIRL problem, a unique bi-policy exists with high probability. 
\section{Linear d-MIRL} \label{section_4}
In \cite{Reddy2012}, Reddy \emph{et al.} consider a decentralized version of MIRL, assuming that the reward of any agent in the multi-agent system only depends on state. Here we extend their approach to a two-person zero-sum MIRL problem in which each agent's reward depends on state and the actions of both agents. As before, let $r$ denote player 1's reward vector. 
\par
In \cite{Reddy2012}, the assumption is made that in a multi-agent system all agents reach a \emph{Markov Perfect Equilibrium} (MPE).  This implies that, for all $s \in \mathcal{S}$ and all $i\in \mathcal{A}_1$,
\begin{equation*}
Q_{\pi}\left ( s \right )\geqslant Q_{\pi|a^1=i}\left ( s \right ).
\end{equation*}
\par
In \cite{Reddy2012}, rewards are selected to maximize the difference between the $Q$ value of the observed policy and those of pure strategies, which is analogous to the classical approach to single-agent IRL given in \cite{Ng2000}.  For our notation, the equivalent problem for agent 1 is the following linear program: 
\begin{equation*} 
\begin{aligned}
\textrm{maximize:}  \quad
& \sum_{s=1}^{N}\mbox{min }_{i\in \mathcal{A}_1} \left ( \tilde{r}_{\pi}\left ( s \right )-\tilde{r}_{\pi|a^1=i}\left ( s \right ) \right ) \\
& +\gamma \left ( G_{\pi}\left ( s \right )- G_{\pi|a^1=i}\left ( s \right ) \right )\left ( I-\gamma G_{\pi}\right )^{-1}B_{\pi}r   \\
& -\lambda \left \| r \right \|_1 \\
\textrm{subject to:} \quad
&\left ( B_{{\pi}^2|a^1=i}-B_{{\pi}} \right )D_{{\pi}}{r}\leq {0},\\
\end{aligned}
\end{equation*}
where $\lambda$ is an adjustable penalty coefficient for having too
many non-zero values in the reward vector. 
\section{Bayesian IRL} \label{section_5}
In this section, we will model the two-person zero-sum multi-agent inverse problem as an IRL problem, by focusing on one agent, which can be called the \emph{agent of interest} and regarding the other agent as part of the inadaptive environment. We extend the BIRL approach developed in \cite{Qiao2011}, which is only applicable to state-dependent reward recovery, to our case where the reward depends on both state and the action of the agent of interest. Note that the reward we want to recover is $r\left ( s,a^1 \right )$ instead of $r\left ( s,a^1, a^2 \right )$, or $r\left ( s,a^1, j \right )=r\left ( s,a^1 \right )$ for all $j \in \mathcal{A}_2$. Although we now turn to the MDP framework, the terminology and notation introduced in Section \ref{section_2} will be used here, unless otherwise specified. 
\par
In \cite{Qiao2011}, rewards are selected to maximize the posterior of the observed state-action pairs given a reward vector $r$, with the likelihood being $1$ if the observed actions are optimal and $0$ otherwise for $r$. For our notation, the equivalent problem for agent 1 is the following linear program: 
\begin{equation}\label{IRL_complex_program}
\begin{aligned}
\textrm{minimize:}  \quad
& \frac{1}{2}\left ( {r-{\mu_r}} \right )^T {\Sigma_r}^{-1} \left ( {r-{\mu_r}} \right )   \\
\textrm{subject to:} \quad
& \left ( F^{\pi^1}_{a^1=i} - C_{a^1=i}  \right ) r \geqslant 0,
\end{aligned}
\end{equation}
for all $i \in\mathcal{A}_1$, where
\begin{equation*}
F^{\pi^1}_{a^1=i} = \left [ \gamma\left ( G_{\pi} - G_{{\pi}^2|a^1=i} \right )\left ( I - \gamma G_{\pi}\right )^{-1} + I  \right ]C_{\pi^1},
\end{equation*}
and where $C_{\pi^1}$ is a $N \times NM$ sparse matrix constructed from $\pi^1$, whose $i$th row is,
\begin{equation*}
\left [ \underbrace{0, \cdots, \pi^1\left ( i, 1 \right ), \cdots ,0}_{N},\underbrace{\cdots}_{\left ( M-2 \right )N} , \underbrace{0, \cdots,\pi^1\left ( i, M \right ), \cdots ,0}_{N}\right ],
\end{equation*}
and $C_{a^1=i}$ is conceptually similar to $C_{\pi^1}$, except for being constructed from a pure policy.
\par
In the above formulation, $\mu_r$ is the mean of the unknown reward vector as a prior, and $\Sigma_r$ is its covariance matrix. Note here we use the notation introduced in Section \ref{subsec:prior}.
\section{Numerical Example}\label{section_6}
In this section, we illustrate the BMIRL method developed in the previous sections on a two-player stochastic game modeled on soccer, and compare results with those obtained from d-MIRL and IRL. Though styled after soccer abstractions in \cite{Littman1994}, the game considered here is richer in that it models an action {\em shoot}, which is a direct attempt to score through a ball kick. 
\par
\subsection{Game and Model}
\begin{figure}[h]
  \centering
  \includegraphics[width=3in]{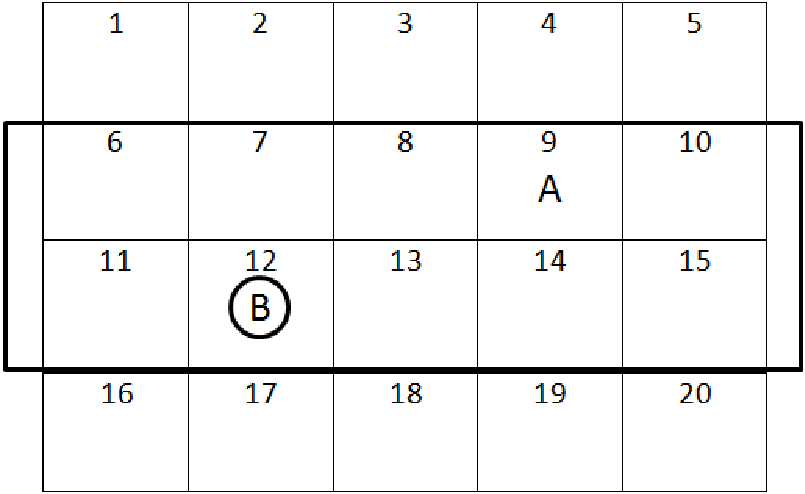}\\
  \caption{Soccer game: initial board}\label{soccer_reset}
\end{figure}

The game is played on a $4 \times 5$ grid as depicted in Figure \ref{soccer_reset}. We use A and B to denote two players, and the circle in the figures to represent the ball. Each player can either stay unmoved or move to one of its neighborhood squares by taking one of 5 actions in each turn: \emph{N} (north), \emph{S} (south), \emph{E} (east), \emph{W} (west), and \emph{stand}. If both players land on the same square in the same time period, the ball is exchanged between the two players with some probability. In addition, the player who has the ball can {\em shoot}, which is to kick the ball toward their opponent's goal, with a \emph{probability of successful shot} (PSS) distribution shown in Table \ref{table:original PSS}. A shot can be taken from any field position, and the PSS is independent of opponent's position. It is worth noting that the PSS at one spot is the probability that the agent believes she would make a successful shot if she kicked the ball at that spot, rather than the actual probability of success she achieves during the play. Otherwise the PSS can be statistically calculated easily through observations once we have inferred the goal area by applying an appropriate MIRL approach. 
\par
In the game setting, both players act simultaneously in each time period. Player A attempts to score by reaching with the ball or shooting the ball into squares 6 or 11, and player B attempts to score by reaching with the ball or shooting the ball into squares 10 or 15. Once a point is scored or a shooting is missed, the players take the positions shown in Figure \ref{soccer_reset} and ball possession is assigned randomly.  
\par
As a third-party observer, we have very limited knowledge about the game they play. We know that this is a zero-sum game. We also know that both players aim to score points by taking or kicking the ball to somewhere in the field. Assume that we watch their playing sufficiently long so that we can statistically calculate their complete policies and their ball exchange rate $\beta = 0.6$ with a perfect accuracy. We will infer which squares each player must reach in order to score a point (the goal squares), as well as the PSS of each player, by means of recovering their reward vector. For example, the PSS of A in position $pos$ ($pos = 1,2,\cdots ,20$) equals the corresponding reward value because 
\begin{eqnarray*}
r\left ( s, a^1 = \mbox{kick}, a^2 \right ) &= & 0\times \left ( 1-PSS^1_{pos} \right )+1\times PSS^1_{pos}\\
&=&PSS^1_{pos},
\end{eqnarray*}
where $s$ is the state where A's position is $pos$. 
There are in total $800$ states in this model, corresponding to the positions of the players and ball possession. Since each player has 6 different actions to choose, each one has a reward vector with a length of $800 \times 6 \times 6 = 28800$. Both players aim to maximize their own total expected points scored, subject to discount factor of $\gamma = 0.9$. 
\par
\begin{table}[h]
\centering 
\begin{tabular}{c c c c c c}  
\hline\hline 
 & PSS = 0.7 & PSS = 0.5 & PSS = 0.3 & PSS = 0.1 & PSS = 0 \\  
A & 1, 7, 12, 16 & 2, 8, 13, 17 & 3, 9, 14, 18 & 4, 10, 15, 19 & 5, 20 \\ 
\hline\hline 
 & PSS = 0.7 & PSS = 0.5 & PSS = 0.3 & PSS = 0.1 & PSS = 0 \\  
B & 5, 9, 14, 20 & 4, 8, 13, 19 & 3, 7, 12, 18 & 2, 6, 11, 17 & 1, 16 \\
\hline 
\end{tabular}
\caption{Original PSS distribution of each player} 
\label{table:original PSS} 
\end{table}
It is worth mentioning that in the simulations done in Section \ref{section_7} the PSS of the two agents happens to be symmetric.  As there is some possibility this structure might give rise to confusion with the negative symmetry property of rewards, note that reward symmetry is due to the precondition of zero-sum and is unrelated to the PSS distributions of the agents. The experiments could be performed with arbitrary PSS and ball exchange probabilities.    
\subsection{Specification of Prior Information} \label{subsec:prior}
Recall that the MIRL optimization program requires the specification of two Gaussian prior parameters for A, the mean of the rewards vector ${\mu_r}$ and the covariance matrix $\Sigma_{{r}}$. 
Below we define a concept of strength for prior information that can be expressed independently in the mean and covariance matrix.
Later subsections focus on the impact of different priors on the quality of learned rewards.
\subsubsection{Mean of the Prior}
We will use three types of mean reward vectors, namely \emph{weak mean}, \emph{median mean} and \emph{strong mean}, respectively. Note that since this
is a zero-sum game, the rewards assigned to B are the negatives of these rewards assigned to A.
\begin{itemize}
\item \emph{Weak Mean}: we assign $0.8$ point to player A in every state where A has possession of the ball and $-0.8$ point in every state where player B has possession of the ball;
\item \emph{Median Mean}: guessing that A's goal might be among the rightmost squares, or squares $5$, $10$, $15$ and $20$, and symmetrically, B's goal might be among the leftmost squares, or squares $1$, $6$, $11$ and $16$, we assign $1$ point to A whenever A has the ball and is in the four leftmost squares, and $-1$ point to A whenever B has the ball and is in four rightmost squares. Also, when A has the ball and takes a shot, no matter where she is, we assign $0.5$ point to A. Similarly, we assign $-0.5$ point to A when B has the ball and takes a shot. Otherwise, no points will be assigned to A.
\item \emph{Strong Mean}: we have a foresight to predict where the goals are for both players, but cannot make a good guess of their PSS distributions. So comparing to \emph{median mean}, the only difference is that now the potential goal area includes only $2$ squares (square $6$ and $11$ for A and square $10$ and $15$ for B), rather than $4$ squares, for both players.
\end{itemize}
\par
\subsubsection{Covariance Matrix}
The covariance matrix of the reward vector encodes our belief of the structure of the prior. Based off of our knowledge of this soccer game, we can develop two types of covariance matrices. 
\begin{itemize}
\item \emph{Weak Covariance Matrix}: an identity matrix, indicating that the reward vector is assumed independently distributed. This is a universal covariance matrix suitable for those MIRL problems in which we neither have knowledge of  the structure of unknowns, nor want to make a guess.  
\item \emph{Strong Covariance Matrix}: a more complex matrix encapsulating some internal information of the reward structure subject to our following beliefs.   
\begin{enumerate}
\item When A has the ball and takes a shot, the PSS depends only on A' s position in the field; likewise for B. 
\item In any state, the reward for A for any non-{\em shoot} action is a state-dependent constant; likewise for B.  
\end{enumerate}
\end{itemize}
Note that the strong covariance matrix can be constructed from the correlation matrix, by assuming that the standard deviation of each random variable in the unknown reward vector is the same. In order to avoid singularity, we will add a small perturbation $\alpha$ to the diagonal of the covariance matrix. 
\par
\subsection{Results Evaluation Metric}
To evaluate a recovered result, we simply compute its \emph{average reward distance} (ARD), which is the average \emph{Euclidean distance} from the true rewards as follows: 		
\begin{equation}
\begin{aligned}
\mbox{ARD} = &\left \{\frac{1}{2NM^2}\left [ \left ( {r}_{\mbox{rec}}^1-{r}^1 \right )^T\left ( {r}_{\mbox{rec}}^1-{r}^1 \right ) \right. \right.\\
&\left. \left. + \left ( {r}_{\mbox{rec}}^2-{r}^2 \right )^T\left ( {r}_{\mbox{rec}}^2-{r}^2 \right ) \right ] \right \}^{1/2},	
\end{aligned}
\end{equation}
where the $NM^2 \times 1$ column vector ${r}_{\mbox{rec}}^k$ and ${r}^k$ denote the recovered and original reward of player $k$. Obviously, the smaller the ARD is, the more accurate the result is.
\par
If only the players' PSS distributions are of interest, a similar version of the evaluation metric, termed \emph{Average PSS Distance} (APD) can be defined as
\begin{equation}
\mbox{APD} = \left \{ \frac{1}{40}\left [ \sum_{i=1}^{20}\left ( {\theta}^1_{\mbox{rec}}\left ( i \right ) - {\theta}^1_{0}\left ( i \right ) \right )^2+ \left ( {\theta}^2_{\mbox{rec}}\left ( i \right ) - {\theta}^2_{0}\left ( i \right ) \right )^2 \right ] \right \}^{1/2},
\end{equation}
where the $20 \times 1$ column vector ${\theta}^k_{\mbox{rec}}$ and ${\theta}^k_{0}$ denote the recovered and original PSS of player $k$, respectively.
\begin{figure*}[!htb]
\centering
\subfloat[Inferred Rewards]{
\label{fig:result_weakmean_weakcov}
\begin{minipage}[t]{0.45\textwidth}
\centering
\includegraphics[width=2.8in]{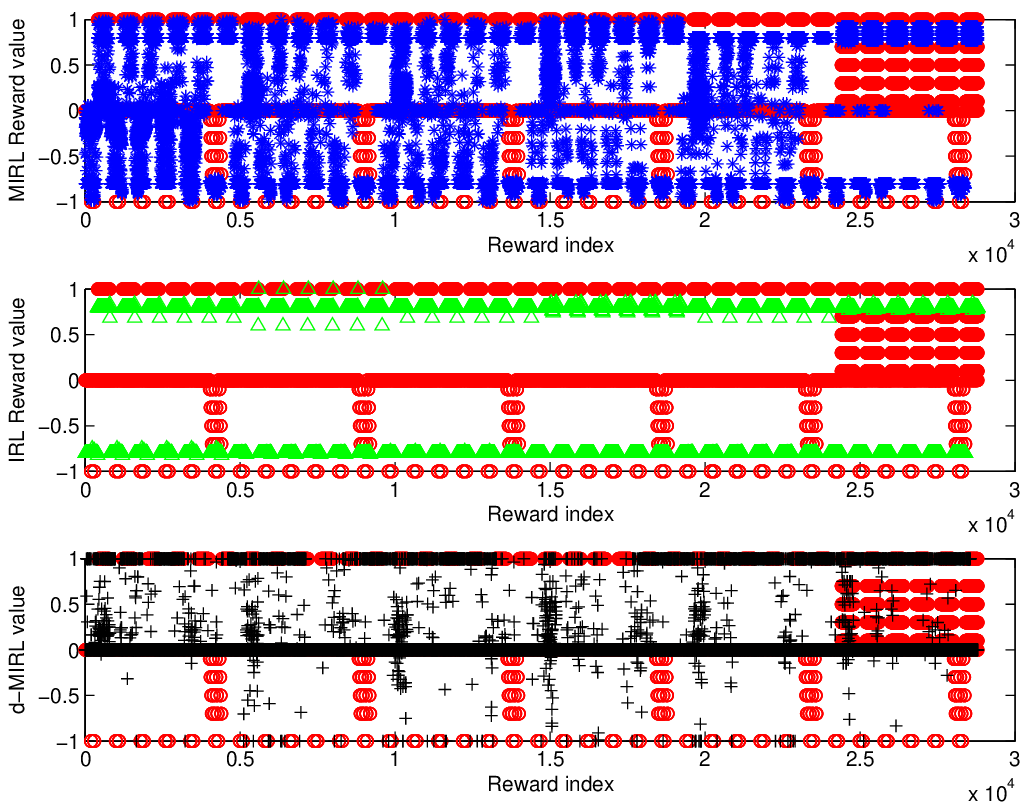}
\end{minipage}
}
\subfloat[Inferred PSS]{
\label{fig:pss_weakmean_weakcov}
\begin{minipage}[t]{0.45\textwidth}
\centering
\includegraphics[width=2.8in]{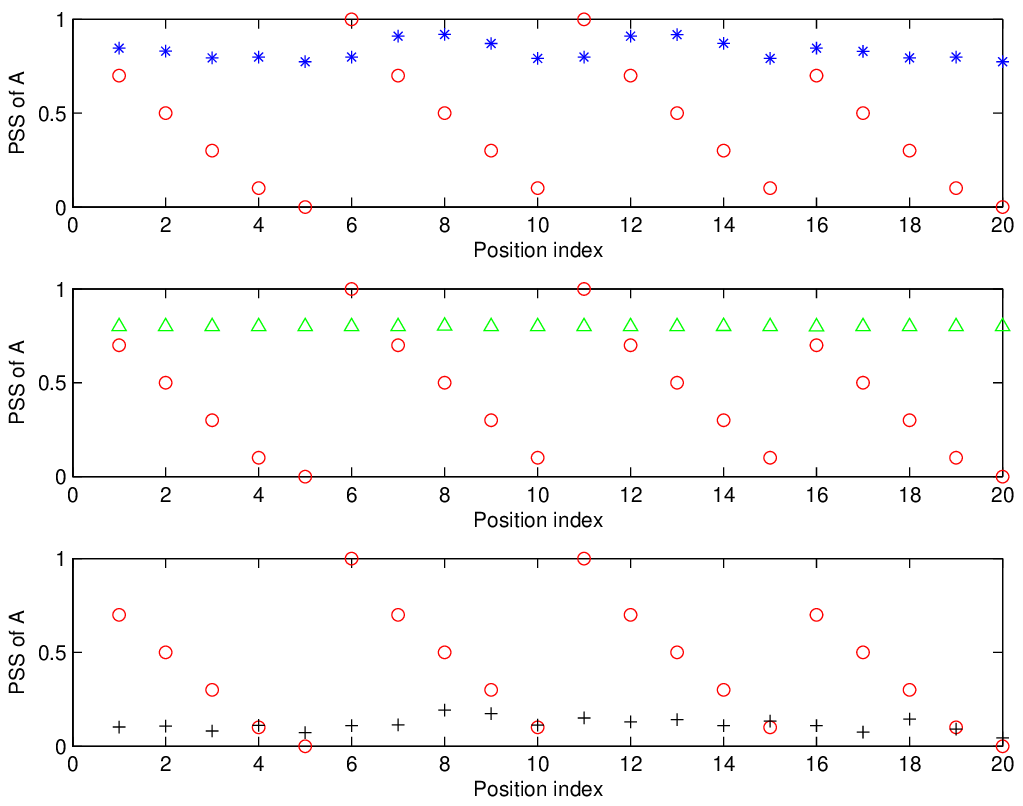}
\end{minipage}
}
\caption{Inferred rewards and PSS: weak mean \& weak covariance}
\label{fig:weakmean_weakcov}
\end{figure*}

\begin{figure*}[!htb]
\centering
\subfloat[Inferred Rewards]{
\label{fig:result_weakmean_strongcov}
\begin{minipage}[t]{0.45\textwidth}
\centering
\includegraphics[width=2.8in]{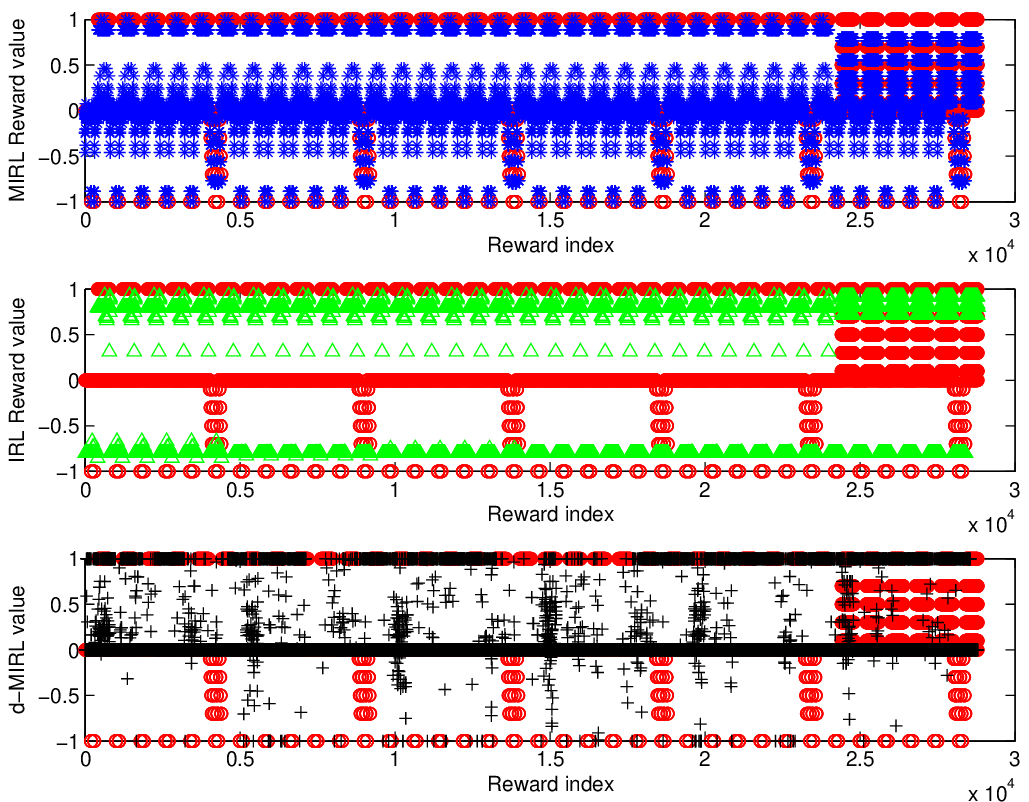}
\end{minipage}
}
\subfloat[Inferred PSS]{
\label{fig:pss_weakmean_strongcov}
\begin{minipage}[t]{0.45\textwidth}
\centering
\includegraphics[width=2.8in]{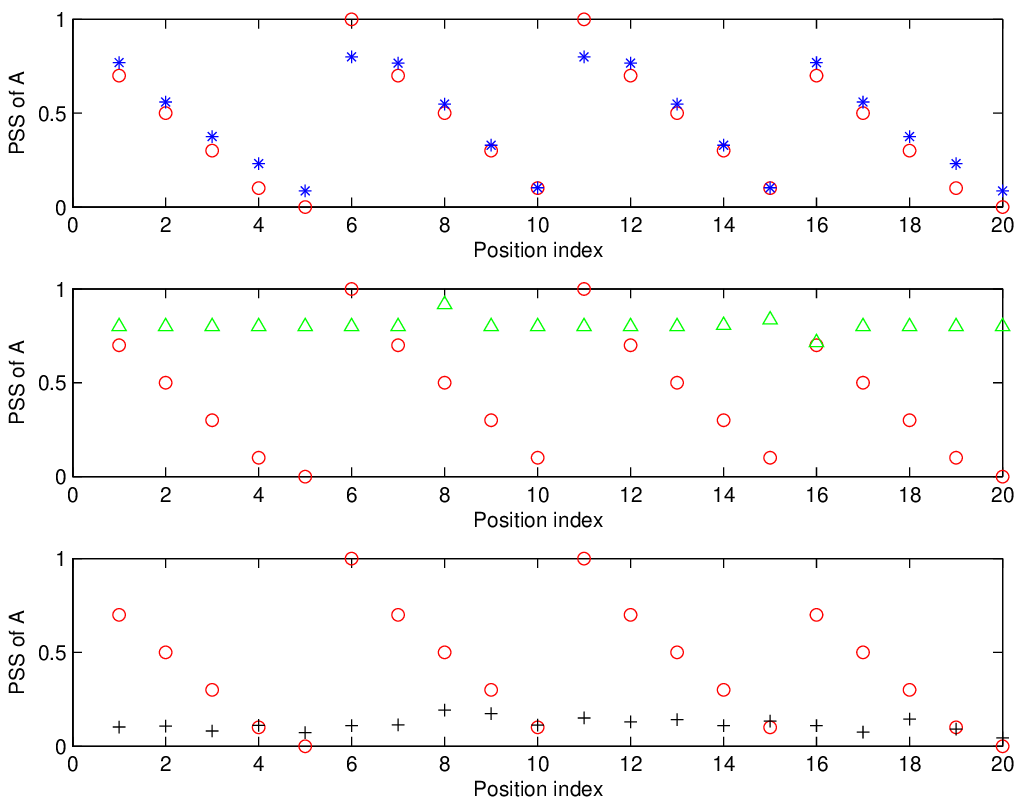}
\end{minipage}
}
\caption{Inferred rewards and PSS: weak mean \& strong covariance}
\label{fig:weakmean_strongcov}
\end{figure*}

\begin{figure*}[!htb]
\centering
\subfloat[Inferred Rewards]{
\label{fig:result_medianmean_weakcov}
\begin{minipage}[t]{0.45\textwidth}
\centering
\includegraphics[width=2.8in]{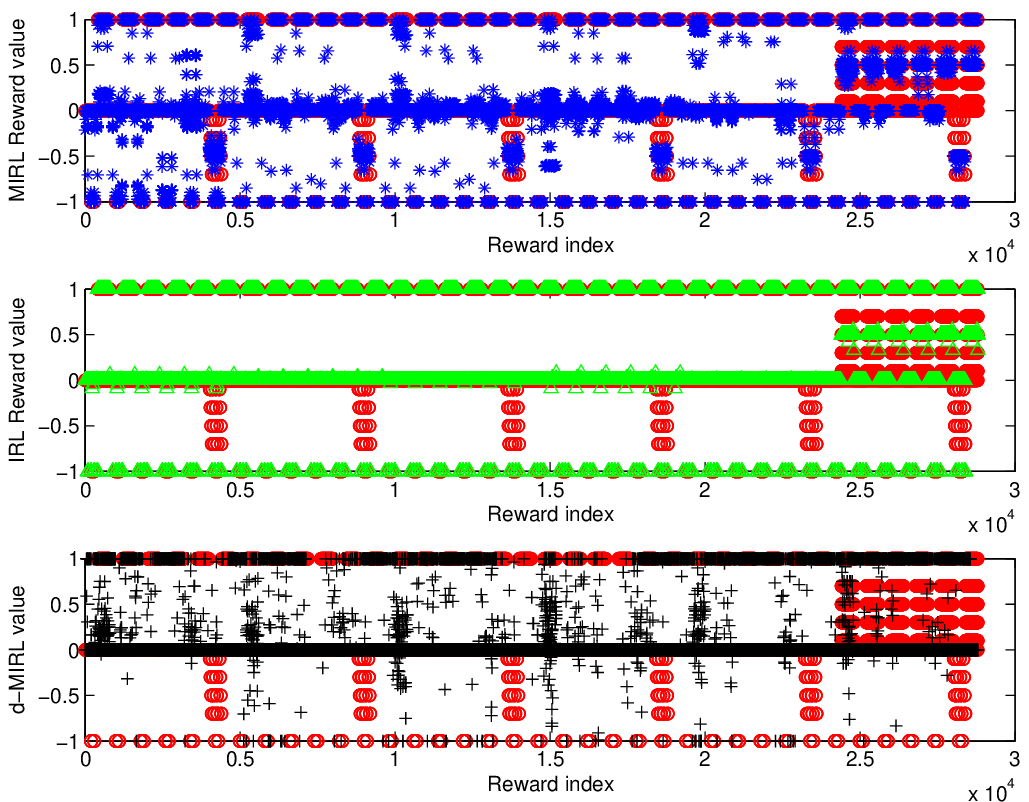}
\end{minipage}
}
\subfloat[Inferred PSS]{
\label{fig:pss_medianmean_weakcov}
\begin{minipage}[t]{0.45\textwidth}
\centering
\includegraphics[width=2.8in]{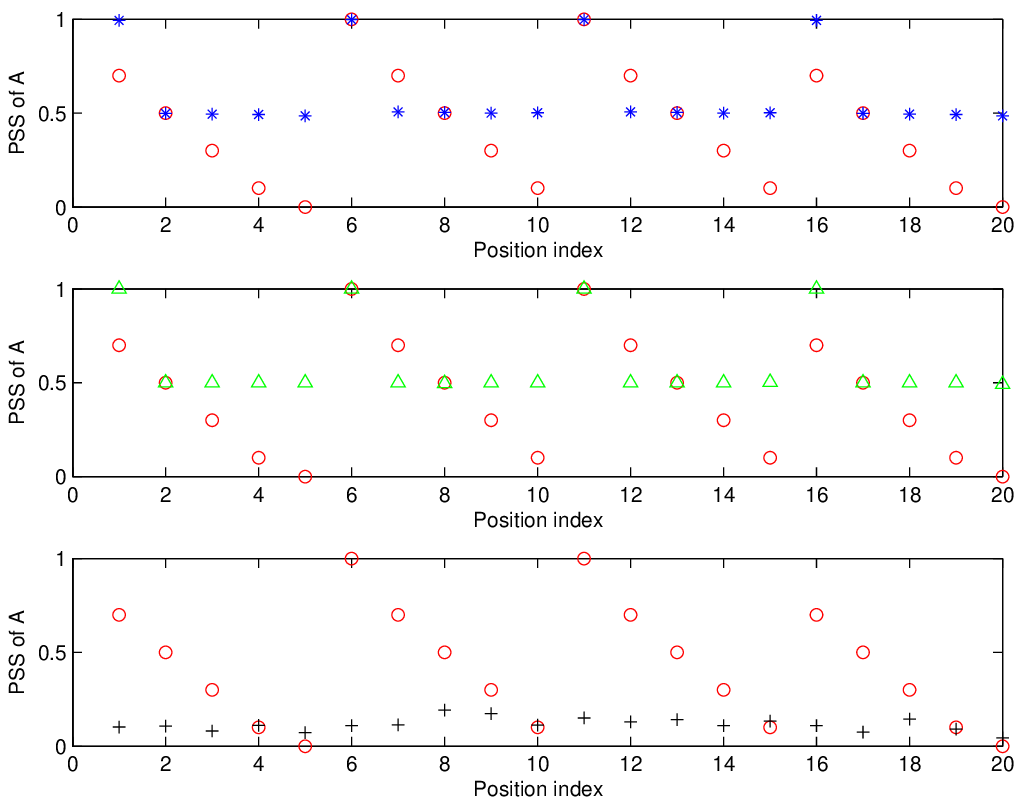}
\end{minipage}
}
\caption{Inferred rewards and PSS: median mean \& weak covariance}
\label{fig:medianmean_weakcov}
\end{figure*}

\begin{figure*}[!htb]
\centering
\subfloat[Inferred Rewards]{
\label{fig:result_medianmean_strongcov}
\begin{minipage}[t]{0.45\textwidth}
\centering
\includegraphics[width=2.8in]{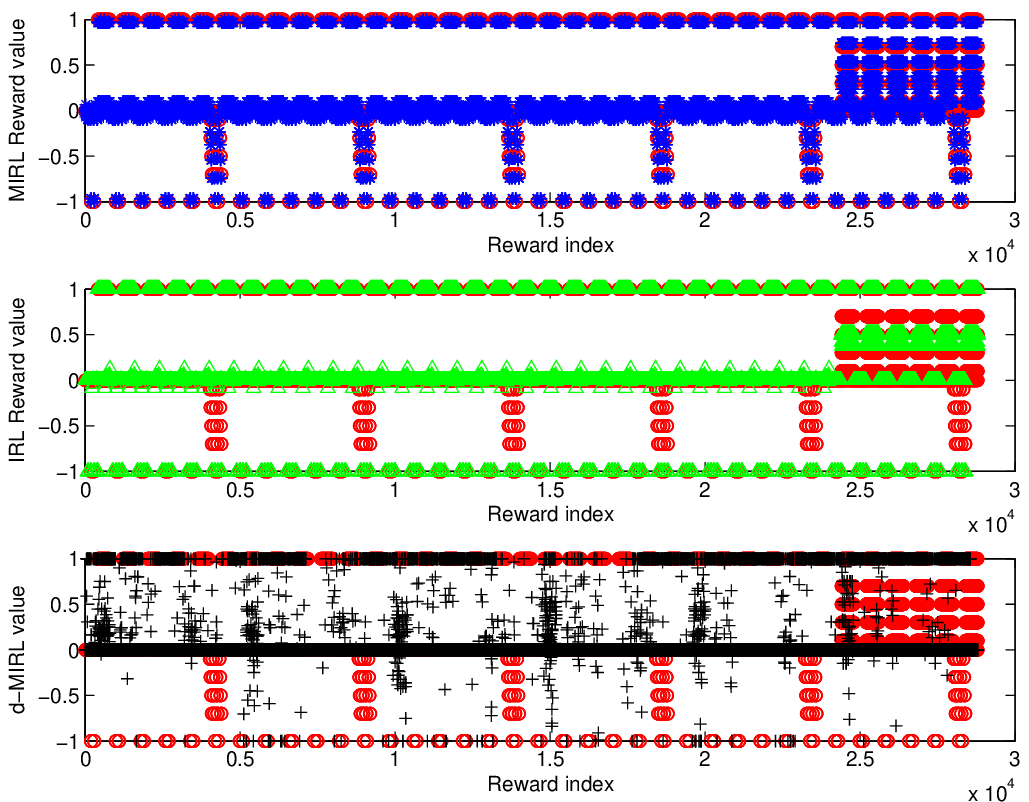}
\end{minipage}
}
\subfloat[Inferred PSS]{
\label{fig:pss_medianmean_strongcov}
\begin{minipage}[t]{0.45\textwidth}
\centering
\includegraphics[width=2.8in]{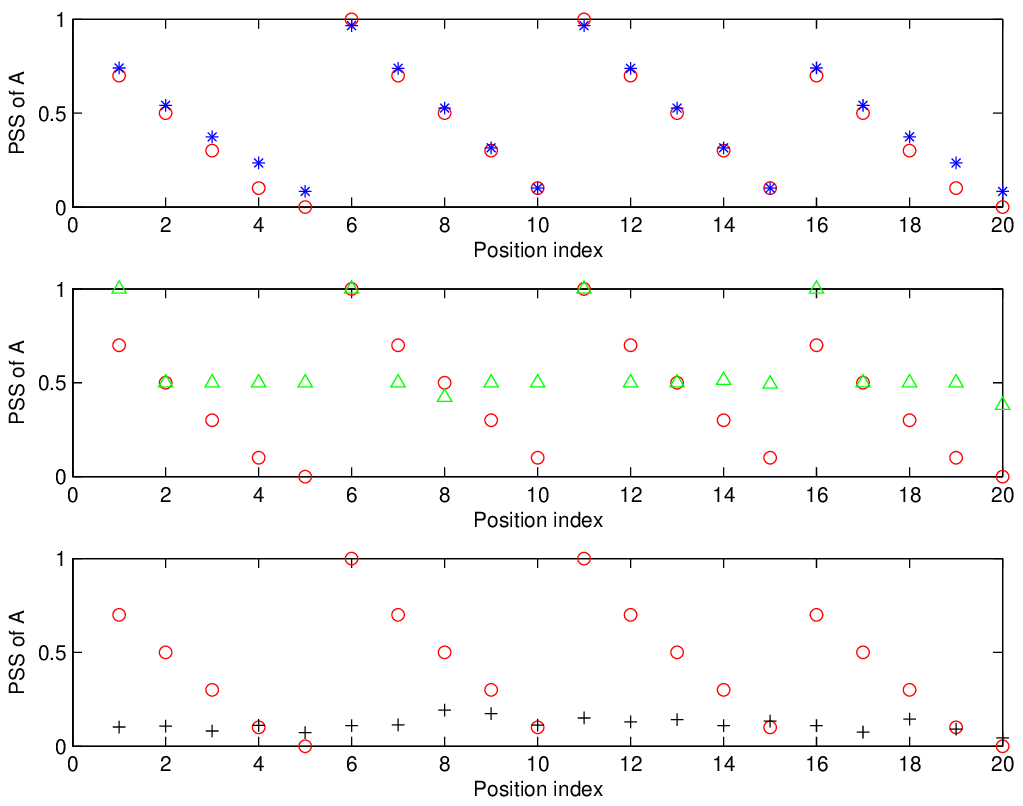}
\end{minipage}
}
\caption{Inferred rewards and PSS: median mean \& strong covariance}
\label{fig:medianmean_strongcov}
\end{figure*}

\begin{figure*}[!htb]
\centering
\subfloat[Inferred Rewards]{
\label{fig:result_strongmean_weakcov}
\begin{minipage}[t]{0.45\textwidth}
\centering
\includegraphics[width=2.8in]{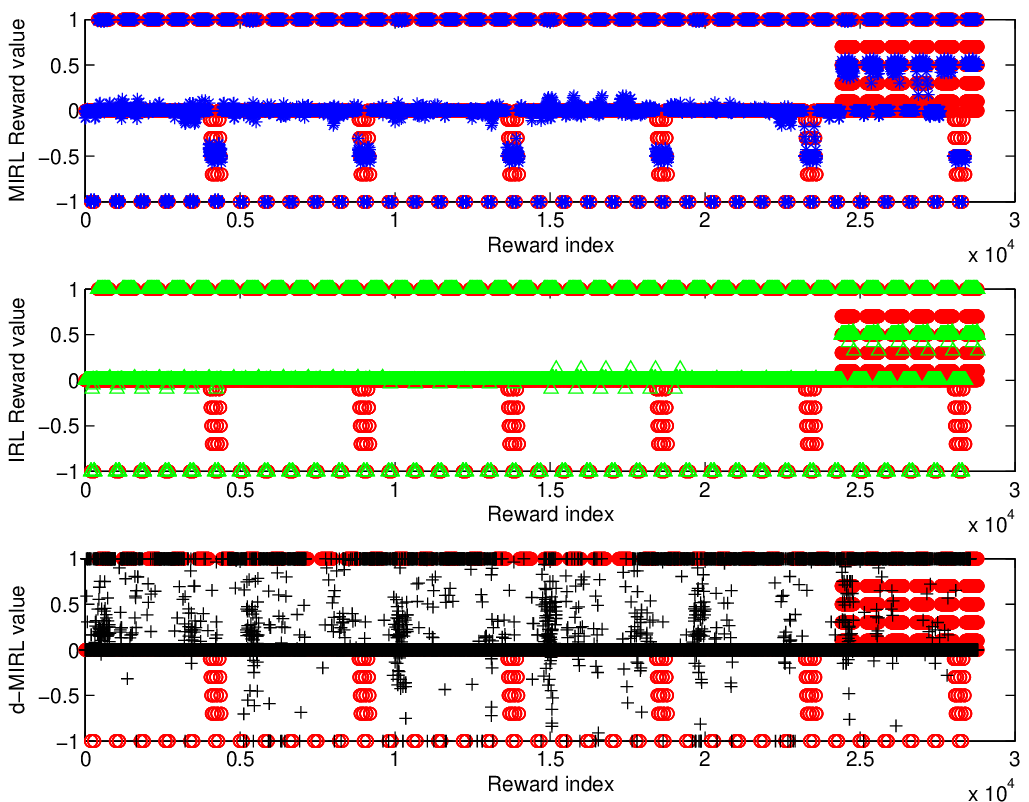}
\end{minipage}
}
\subfloat[Inferred PSS]{
\label{fig:pss_strongmean_weakcov}
\begin{minipage}[t]{0.45\textwidth}
\centering
\includegraphics[width=2.8in]{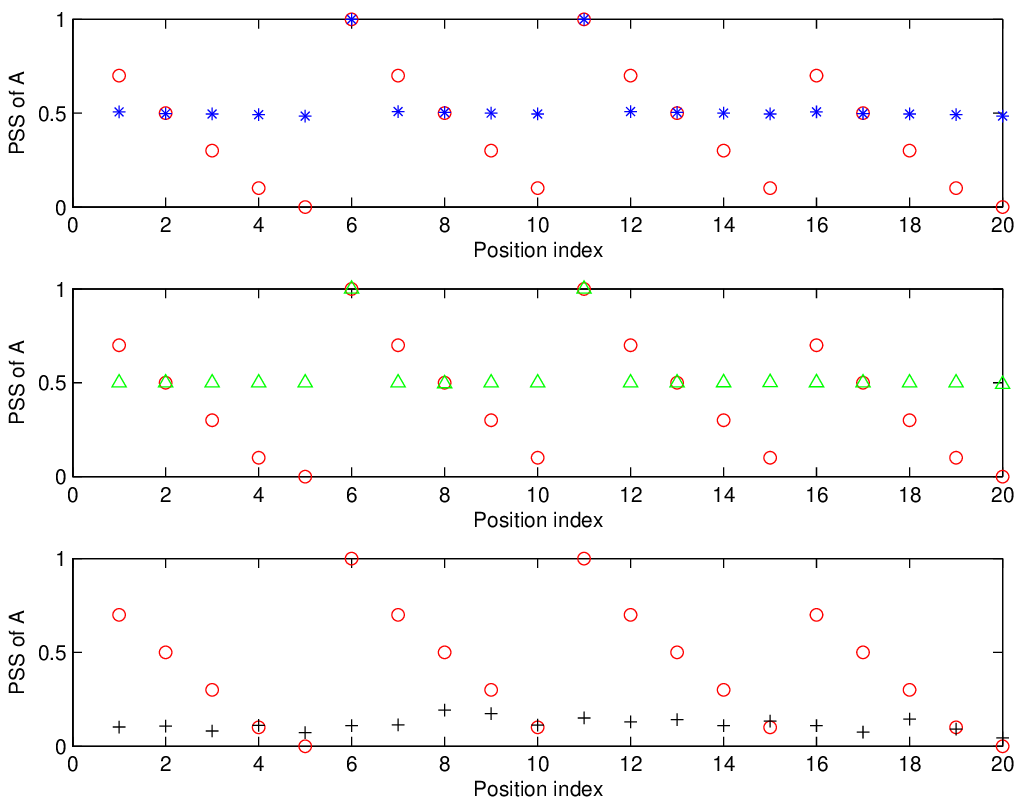}
\end{minipage}
}
\caption{Inferred rewards and PSS: strong mean \& weak covariance}
\label{fig:strongmean_weakcov}
\end{figure*}

\begin{figure*}[!htb]
\centering
\subfloat[Inferred Rewards]{
\label{fig:result_strongmean_strongcov}
\begin{minipage}[t]{0.45\textwidth}
\centering
\includegraphics[width=2.8in]{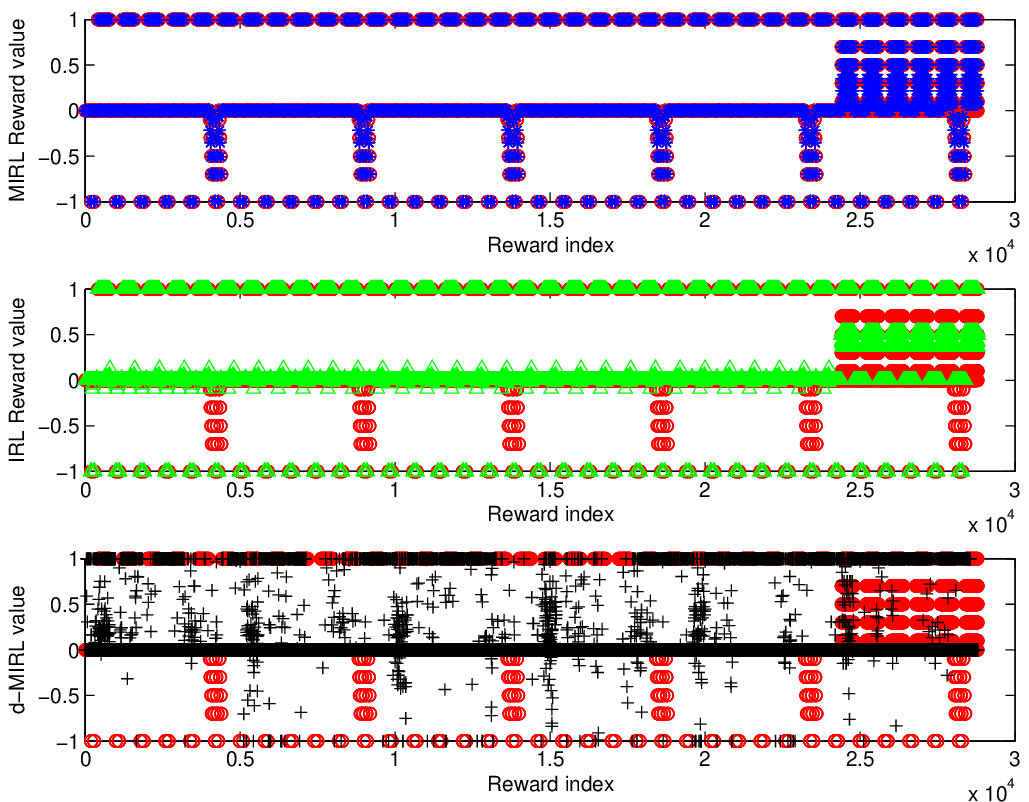}
\end{minipage}
}
\subfloat[Inferred PSS]{
\label{fig:pss_strongmean_strongcov}
\begin{minipage}[t]{0.45\textwidth}
\centering
\includegraphics[width=2.8in]{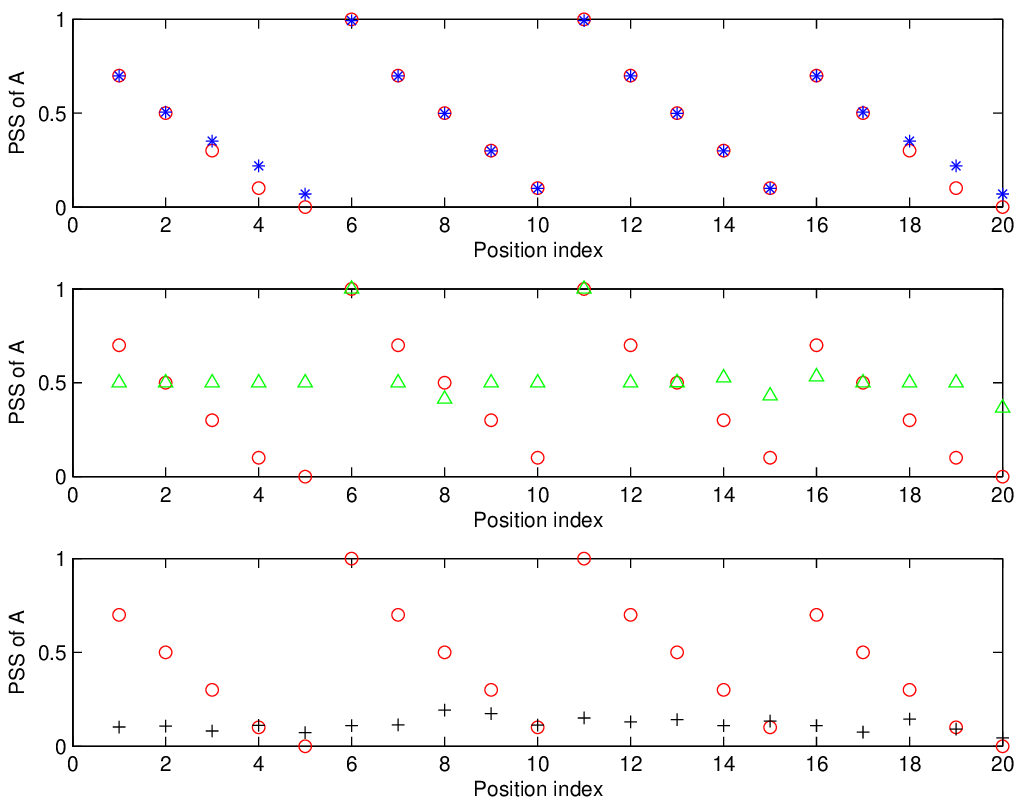}
\end{minipage}
}
\caption{Inferred rewards and PSS: strong mean \& strong covariance}
\label{fig:strongmean_strongcov}
\end{figure*}
\par
\subsection{Results}
Experiments were performed on 6 different priors formed by combining 3 different means and 2 different covariance matrices. A pertubation $\alpha = 10^{-4}$ was used in the construction of the strong covariance matrices. In all cases, the bi-policy followed by the players (the observed input to MIRL) was computed iteratively from Shapley's Theorem, discussed in Section \ref{sec:zerosum}. Experiments on Bayesian IRL (we can also specify 6 different priors similar to those introduced in Section \ref{subsec:prior}) and d-MIRL were also carried out. Note that the reward vector recovered from IRL can be extended to a MIRL reward vector by letting $r\left ( s, a^1, j \right ) = r\left ( s, a^1\right )$ for all $j \in \mathcal{A}_2$.
\par
\begin{figure}[htbp]
\centering
\begin{minipage}[t]{0.45\textwidth}
\centering
\includegraphics[width=2.8in]{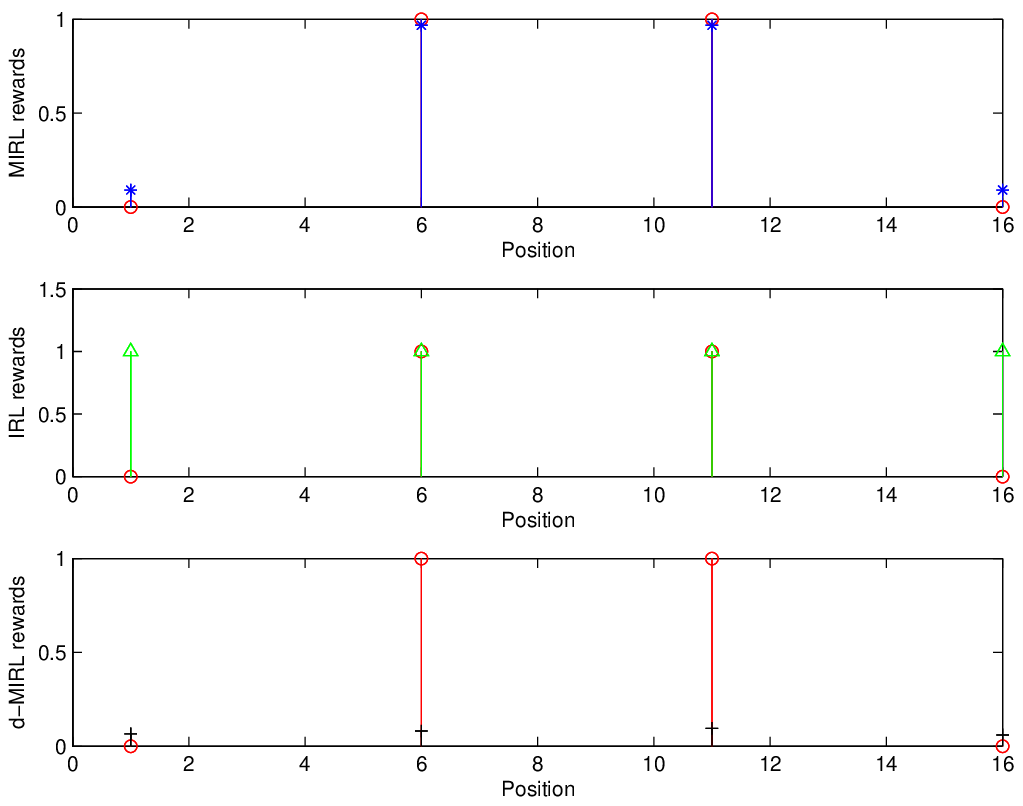}
\caption{Goal recovery in Case 4}
\label{fig:where_goals}
\end{minipage}
\begin{minipage}[t]{0.50\textwidth}
\centering
\includegraphics[width=2.8in]{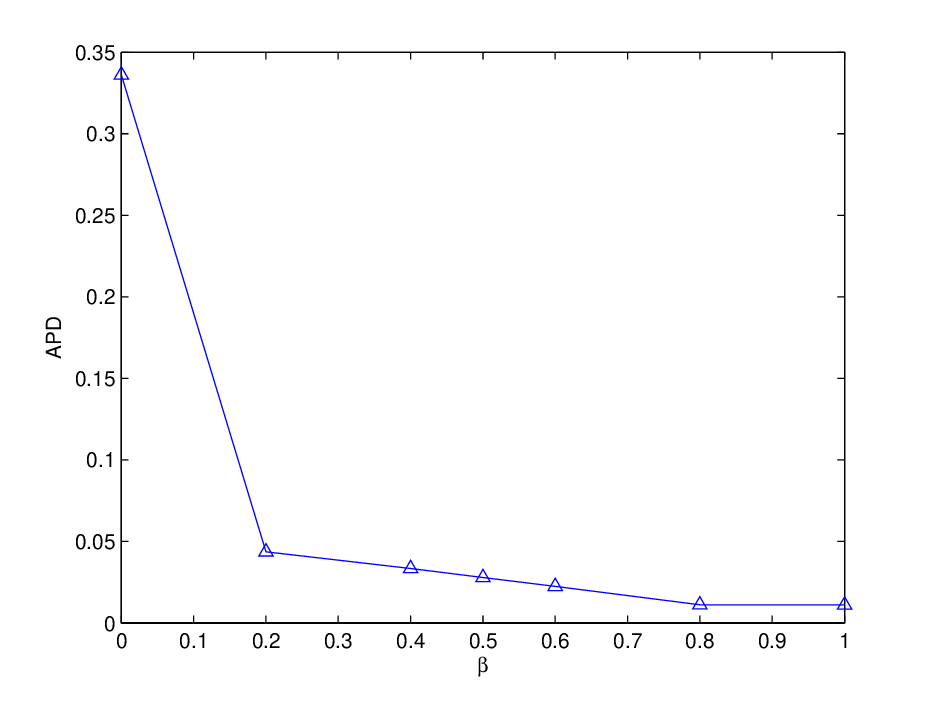}
\caption{APD with $\beta$ changing in Case 6}
\label{fig:static_APD_strongprior_strongcov}
\end{minipage}
\end{figure}

Results are shown in Figures \ref{fig:weakmean_weakcov}-\ref{fig:strongmean_strongcov}. Take Figure \ref{fig:weakmean_strongcov} as an example. Recall that we aim to recover 28800 reward values. In each subfigure in (a), the x-axis represents the reward value index (from 1 to 22800) and the y-axis denotes the reward value. The inferred rewards of BMIRL, BIRL and d-MIRL are shown in blue stars, green triangles and black crosses, respectively, with the benchmark ground truth drawn in red circles in each subfigure. The right three subfigures in (b) show the results of A's PSSs corresponding to each case. Note that although no shots will be taken at goal positions, for convenience, we set $PSS=1$ for each player in their goal positions. Table \ref{table: PSS accuracy of the recovered BMIRL rewards} sorts each experiment with a case number, maps each case to a figure and computes the corresponding APD of the BMIRL rewards. 
\begin{table}[h]
\centering 
\begin{tabular}{c c c  }  
\hline\hline 
  & Weak Covariance & Strong Covariance \\  
\hline 
Weak Mean & Case 1, Figure 2, 0.4535 & Case 3, Figure 4, 0.0671  \\ 
Median Mean & Case 3, Figure 4, 0.2169 & Case 4, Figure 5, 0.0387 \\
Strong Mean & Case 5, Figure 6, 0.2058 & Case 6, Figure 7, 0.0259  \\
\hline 
\end{tabular}
\caption{BMIRL results summary} 
\label{table: PSS accuracy of the recovered BMIRL rewards} 
\end{table}
In Case 4, we are also interested in whether the three methods can recover the actual goals for A. We calculate the average reward A receives when A is in square $1$, $6$, $11$ and $16$. Results are shown in Figure \ref{fig:where_goals}. 
Now focus on the BMIRL method. It is interesting to consider how the ball exchange rate $\beta$ affects the PSS recovery result. We repeat Case 6 by changing $\beta$ from $0$ to $1$, and calculate the APD of the inferred PSS distributions. The result is shown in Figure \ref{fig:static_APD_strongprior_strongcov}.
\par
\subsection{Analysis of Results}
From Figures \ref{fig:weakmean_weakcov}-\ref{fig:strongmean_strongcov} and Table \ref{table: PSS accuracy of the recovered BMIRL rewards}, we can easily come to a conclusion that generally, among the three methods, BMIRL performs much better than the other two. For BMIRL, 
\begin{itemize}
\item The closer the mean is to the actual rewards, the better the quality of learned rewards will be, and likewise for the covariance matrix.
\item The covariance matrix has a greater influence on the quality of learned than does the mean.
\end{itemize}
\par
From Figure \ref{fig:where_goals} we see that  BMIRL  successfully learns the goals for A, while the other two methods fail to do so. Finally, Figure \ref{fig:static_APD_strongprior_strongcov} shows that the smaller the $\beta$ is, the less accurate the recovered PSS will be. The reason is that players are inclined to dribble the ball rather than shoot it toward their opponents' goal when $\beta$ is small, and consequently, observing the strategy of dribbling will not generate constraints that substantially alter the mode of the priors on shooting rewards. For example, when $\beta = 0.2$, the probability of successfully dribbling the ball to the destination for each player is, at worst, $(1-\beta)^4 = 0.407$, which means that a shot will never be taken in positions where the agent's PSS is 0.3 or 0.1. 
\section{Monte Carlo Simulation using Recovered Rewards}\label{section_7}
In the previous section, distance metrics in reward and PSS space are used to evaluate the quality of learned rewards. In this section we measure the reward quality in terms of the quality of the forward solution that would be based on the rewards. IRL is often set in the context of apprenticeship learning, in which learned rewards form the basis for anticipating or mimicking the response of agents to unknown situations. In MIRL, the analogous notion is to use learned rewards as the basis for game play in different environmental settings. In this section, we will simulate a series of games, by letting different agents use different rewards generated from the three methods discussed above and play against each other. Being rational, all agents will employ a minimax policy based off of which is the rewards they learned. Specifically, define the following agents:
\begin{itemize}
\item \emph{A}, which uses true rewards;
\item \emph{B}, which uses BMIRL rewards;
\item \emph{C}, which uses BIRL rewards;
\item \emph{D}, which uses d-MIRL rewards.
\end{itemize}
A full set of agent-to-agent competition then includes the following scenarios:
\begin{itemize}
\item B against A;
\item B against C;
\item B against D.
\end{itemize}
All those games are simulated in three different environment settings, where the the ball exchange rates $\beta$ are $0$, $0.4$ and $1$, respectively. Note that the symmetry of PSS values means that the two agents are equally skillful and are supposed to be an equal in match both of them follow reasonable policies generated from learned rewards.
%
\par
The simulation results are presented in Table \ref{table:simulation_results_B_A}-\ref{table:simulation_results_B_D}. In each table, the first column is the different sets of BMIRL rewards that B employs to develop her minimax policy, where \emph{WM}, \emph{MM}, \emph{SM}, \emph{WC} and \emph{SC} stand for \emph{weak mean}, \emph{median mean}, \emph{strong mean}, \emph{weak covariance matrix} and \emph{strong covariance matrix}, respectively. The remaining columns are the \emph{win or lose} (W/L) outcomes of 10000 rounds of games between B and other agents in cases where $\beta$ being $0$, $0.4$ and $1$. For example, in Table \ref{table:simulation_results_B_A}, 24.69/25.10 means B beats A with probability 24.69\% and loses with probability 25.10\%. It indicates that the remaining 50.21\% rounds end in a tie. A tie occurs when neither player scores a point. For a more clear comparison, we only count those game episodes ending in win-lose outcomes. Each column except for the first presents B's winning percentage. Note that in Table \ref{table:simulation_results_B_C}, since there are also 6 sets of BIRL rewards, comparisons are between corresponding sets, e.g., SM-SC BMIRL vs SM-SC BIRL. 
\begin{table}[ht]
\centering 
\begin{tabular}{l c c c} 
\hline\hline 
Base Rewards &  W/L\% ($\beta = 0.4$) &  W/L\% ($\beta = 1$) &  W/L\% ($\beta = 0$)\\ [0.5ex] 
\hline 
WM \& WC & 0/24.80 & 0/62.53 & 0/50.30\\ 
WM \& SC & 24.69/25.10 & 25.10/25.30 & 50.66/49.34\\ 
MM \& WC & 15.28/25.34 & 14.36/24.69 & 28.44/49.43\\ 
MM \& SC & 24.73/25.03 & 24.12/25.18 & 49.84/50.16\\ 
SM \& WC & 14.85/24.52 & 14.94/25.50 & 49.31/50.69\\ 
SM \& SC & 24.77/25.32 & 24.55/25.43 & 49.84/50.16\\ 
\hline 
\end{tabular}
\caption{B vs A games simulation results } 
\label{table:simulation_results_B_A} 
\end{table}
\begin{table}[ht]
\centering 
\begin{tabular}{l c c c} 
\hline\hline 
Base Rewards &  W/L\% ($\beta = 0.4$) &  W/L\% ($\beta = 1$) &  W/L\% ($\beta = 0$)\\ [0.5ex] 
\hline 
WM \& WC & 0/0 & 13.50/0 & 0/0\\ 
WM \& SC & 23.36/0 & 24.64/0 & 50.29/0\\ 
MM \& WC & 13.55/0 & 15.64/0 & 26.82/0\\ 
MM \& SC & 22.73/6.74 & 25.45/14.80 & 49.55/27.58 \\ 
SM \& WC & 15.82/0 & 14.27/0 & 49.87/0\\ 
SM \& SC & 23.36/0 & 24.64/0 & 50.13/0\\ 
\hline 
\end{tabular}
\caption{B vs C games simulation results } 
\label{table:simulation_results_B_C} 
\end{table}
\begin{table}[ht]
\centering 
\begin{tabular}{l c c c} 
\hline\hline 
Base Rewards &  W/L\% ($\beta = 0.4$) &  W/L\% ($\beta = 1$) &  W/L\% ($\beta = 0$)\\ [0.5ex] 
\hline 
WM \& WC & 0/0 & 0/0 & 0/0\\ 
WM \& SC & 25.52/0 & 26.36/0 & 49.98/0\\ 
MM \& WC & 12.52/0 & 16.75/0 & 50.26/0\\ 
MM \& SC & 24.60/0 & 27.30/0 & 49.20/0 \\ 
SM \& WC & 12.24/0 & 13.26/0 & 49.46/0\\ 
SM \& SC & 25.22/0 & 26.48/0 & 49.90/0\\ 
\hline 
\end{tabular}
\caption{B vs D games simulation results } 
\label{table:simulation_results_B_D} 
\end{table}
\par
\begin{figure}[htbp]
\centering
\includegraphics[width=3.5in]{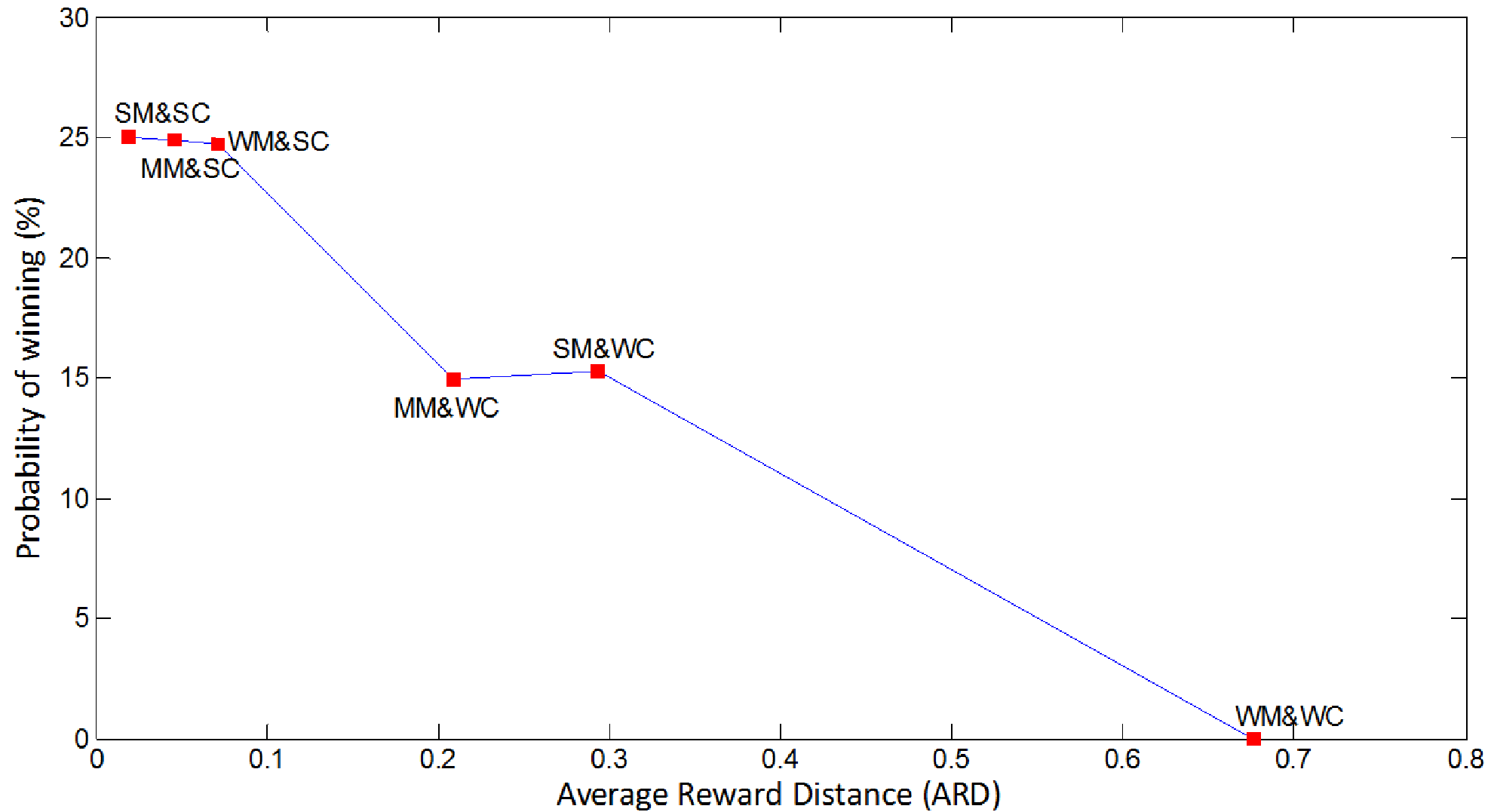}
\caption{Two evaluation metrics comparison}
\label{two_metrics_comparison}
\end{figure}

Let us coin the term \emph{Application Metric} (AM) to refer to B's probability of winning in the soccer example. Table \ref{table:simulation_results_B_A} shows that A outperforms or ties B in general. This result is reasonable because A uses true rewards. In addition, we compare AM with the previous numerical metric ARD in Figure \ref{two_metrics_comparison}. As expected, a larger ARD results in a smaller probability of winning. What is notable is the sudden crash in probability of winning experienced when ARD becomes sufficiently large. Equivalently, the probability of B's winning drops sharply when both the mean and covariance are weak. The implication is that inferring the structure of the unknowns, is much more crucial than inferring their true values. As for the other two methods, Tables  \ref{table:simulation_results_B_C}-\ref{table:simulation_results_B_D} show that B generally outperforms C or D. 
\par
\section{Additional Experiments} \label{section_additional}
Thus far we have demonstrated the performance of our BMIRL algorithm through a numerical experiment. There remain, however, two important questions to address. First, how does our BMIRL approach compare to supervised/semi-supervised learning based policy learning approaches? Second, can we still expect good performance if the game is played on a larger size grid, say, $5*5$?
\par
\begin{figure}[ht]
  \centering
  \includegraphics[width=2.5in]{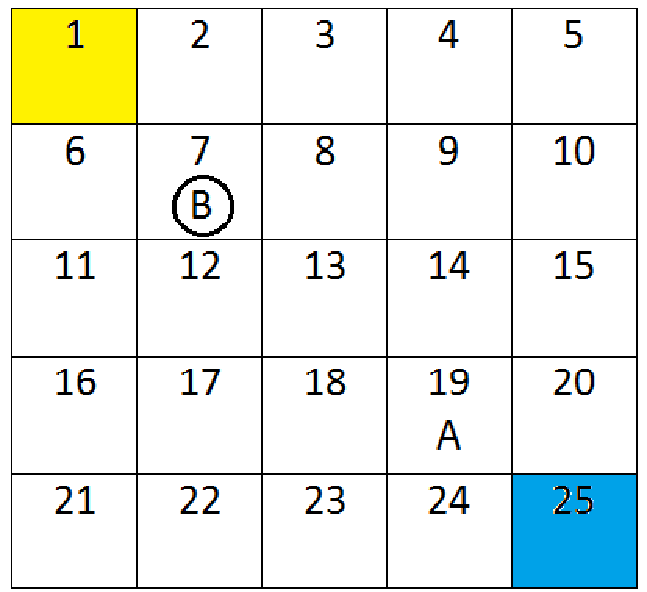}\\
  \caption{Soccer game: 5*5 board}\label{new_game_grid}
\end{figure}

This section is dedicated to addressing these two questions through two more experiments in the context of the soccer game. The first experiment is to use multivariate linear regression to learn a linear relationship between predictors (state and the ball exchange rate) and the response (bi-strategies) and then to infer the response in a new environment. Note that normalization is needed before applying the regression. The second experiment is to re-design the game on a $5*5$ grid, as shown in Figure \ref{new_game_grid}, where A and B's starting positions are 19 and 7, and their goals are 1 and 25, respectively. The PSS distributions are also re-assigned. Other settings and rules of this new game remain as they are in the old one.
\par
Performance evaluations in these two experiments are conducted through Monte-Carlo simulation as in Section \ref{section_7}. Specifically, in the first experiment, we define agent $B_p$ as using policy-learning method and simulate the scenario of $B$ against $B_p$. In the second experiment, we investigate $B_{5*5}$ against $A_{5*5}$, where $B_{5*5}$ and $A_{5*5}$ denote agents using BMIRL rewards and true rewards in the new game, respectively.
\par
The results of the first experiment, presented in Table \ref{table:simulation_results_B_Bp}, show that BMIRL generally outperforms the policy-learning method when a strong covariance matrix is applied in the prior, and generates comparable results with those of the policy-learning method in other cases with the exception of the worst prior condition. In the second experiment, we offer more combinations of mean and covariance as prior information is very critical in the performance of BMIRL. Specifically, we provide one more \emph{median covariance matrix}, denoted as \emph{MC}, subject to our beliefs that: (1) when A has the ball and takes a shot, the PSS depends only on A' s position in the field; and (2) the reward for A for any non-{\em shoot} action is generally strongly correlated. As shown in Table \ref{table:simulation_results_B_A_5_5}, results are similar to those from the experiments reported in Table \ref{table:simulation_results_B_A} and confirm the associated conclusions.	
\begin{table}[ht]
\centering 
\begin{tabular}{l c c c} 
\hline\hline 
Base Rewards &  W/L\% ($\beta = 0.4$) &  W/L\% ($\beta = 1$) &  W/L\% ($\beta = 0$)\\ [0.5ex] 
\hline 
WM \& WC & 0/14.91 & 0/21.30 & 0/36.40\\ 
WM \& SC & 22.54/21.16 & 19.19/18.11 & 47.15/36.45 \\ 
MM \& WC & 20.82/23.38 & 19.70/17.90 & 40.65/36.85\\ 
MM \& SC & 28.89/24.46 & 27.98/16.86 & 49.48/40.08 \\ 
SM \& WC & 19.79/23.61 & 19.52/17.88 & 50.15/35.65\\ 
SM \& SC & 29.04/23.56 & 30.94/20.76 & 50.26/35.44\\ 
\hline 
\end{tabular}
\caption{$B$ vs $B_p$ games simulation results } 
\label{table:simulation_results_B_Bp} 
\end{table}
\begin{table}[ht]
\centering 
\begin{tabular}{l c c c} 
\hline\hline 
Base Rewards &  W/L\% ($\beta = 0.4$) &  W/L\% ($\beta = 1$) &  W/L\% ($\beta = 0$)\\ [0.5ex] 
\hline 
WM \& WC & 20.20/20.60 & 5.07/4.93 & 25.42/49.78\\ 
WM \& MC & 20.90/21.20 & 4.44/4.46 & 24.46/50.34 \\
WM \& SC & 20.12/21.19 & 24.89/25.80 & 43.60/50.24\\ 
MM \& WC & 19.41/18.79 & 4.17/4.23 & 24.19/49.11\\ 
MM \& MC & 20.94/20.86 & 5.32/5.28 & 25.56/49.94 \\
MM \& SC & 21.02/20.60 & 24.27/24.62 & 43.60/49.98\\ 
SM \& WC & 20.02/20.88 & 5.23/5.27 & 25.06/51.34\\ 
SM \& MC & 20.03/20.07 & 4.23/4.37 & 26.14/51.06\\
SM \& SC & 24.81/25.78 & 25.32/24.72 & 49.84/50.16\\ 
\hline 
\end{tabular}
\caption{$B_{5*5}$ vs $A_{5*5}$ games simulation results } 
\label{table:simulation_results_B_A_5_5} 
\end{table}
\par
\section{Conclusions}\label{section_8}
This paper introduces the MIRL problem in the setting of zero-sum stochastic games and presents a solution based on Bayesian inference. Although it seems that MIRL is a natural extension of IRL, it in fact  presents more challenges. Even in simple static games two important distinctions between inverse learning for optimization and inverse learning for games emerge. While the model in this paper assumes that the complete bi-policy of two players is observed, it is more likely that only actions of the individual players are observed. In an optimization setting, since deterministic policies are assumed, strategies can be inferred exactly from finitely many observations of actions. In the case of games, strategies are often mixed, and so strategies cannot be inferred exactly from finitely many observations of the actions taken in each state.  Therefore, we cannot model a player's strategy as an observation as it can be done in IRL. In the setting of games, strategies must be treated as latent variables that are not observed directly, but bridge the gap between reward functions and observable actions.

Though ideally structured,  the numerical examples considered in this section serve to demonstrate the ill-specified nature of the MIRL problem.  Neither BIRL nor d-MIRL perform satisfactorily on the numerical examples. The rationale underlying this phenomenon is that there always exist multiple feasible solutions that are consistent with the observations. It is extremely difficult to select a reward function that is closest to the ground truth without a certain amount of domain knowledge. Our proposed BMIRL approach makes use of domain knowledge expressed as priors on the reward function. That distinction, new to the literature of MIRL methods, is why our Bayesian method is superior to the d-MIRL method in the numerical examples. Fortunately, in many real problems domain knowledge would be available to  observers. 

A principal motivation for the study of MIRL in game settings is that the approach offers insight into how agents will behave if the game environment, rules, or dynamics change.  Such insight may be useful in game design and management, such as balance adjustment.  Effective supervised methods exist for learning policies from observed actions, but policies learned in this fashion do not project into new game environments. The reason is that the optimal policy often changes with environment and hence learning from an old policy may not help to infer a new policy. To see this, consider the abstract soccer game. In Section \ref{section_7}, three additional agents B, C and D come up with their own minimax policies by using rewards learned from three different methods, and compete with A in three different environmental settings: the ball exchange rate $\beta=0, 0.4$ and $1$. Recall that rewards were learned when $\beta=0.6$.  The similarity of two policies, say $p_1$ and $p_2$, can be measured using the \emph{Frobenius distance} $F$, defined as: $F_{p_1,p_2}=\sqrt{\mbox{tr}\left ( \left ( p_1-p_2 \right )\left ( p_1-p_2 \right )' \right )}$.
 Table \ref{table:policy_difference} shows the similarity of player B's policies as a function of $\beta$.  The conclusion to be drawn is that as the environment changes, so does the policy. 
 
\begin{table}[ht]
\centering 
\begin{tabular}{l c c c} 
\hline\hline 
 &  $\beta = 0.4$ &  $\beta = 1$ &  $\beta = 0$\\ [0.5ex] 
$F_{\beta, 0.6}$ & 5.71 & 8.53 & 20.49\\ 
\hline 
\end{tabular}
\caption{Policy difference w.r.t. $\beta$} 
\label{table:policy_difference} 
\end{table}

\section{Future Work}\label{section_9}
Several directions are suggested for future work. First, generative models are needed for more general scenarios, such as those where a full bi-policy is unobservable. Such scenarios have been discussed in \cite{Reddy2012}, under the assumption that the agents have reached a minimax equilibrium. There is room for methods that can incorporate domain knowledge and informative priors. Second, an extension of our method to those MIRL problems where the state transition matrix is difficult to obtain or estimate is needed. The approach in this paper requires the state transition matrix  to be known. It is well known, however, that RL problems can be addressed without knowing the state transition matrix, and this is true for IRL as well \cite{Levine2011}.  An extension of MIRL to such setting may be worthwhile. Third, an extension is needed  to the $n$-player, general-sum case. This will likely be challenging, because in general-sum games, multiple equilibria may be associated with different game values. Moreover, the specific assumptions imposed on equilibrium selection will affect the nature of the reward functions recovered from an inverse learning procedure. A good starting point for study of general sum games would be multi-player, stochastic iterated Prisoners' Dilemma, as the MIRL perspective might allow the interpretation of learned rewards in terms of the dynamics of strategy evolution.


\section*{Acknowledgment}
This work was partially supported by Science Applications International Corporation (SAIC) through the Research Scholars Fellowship Program.
\ifCLASSOPTIONcaptionsoff
  \newpage
\fi



%
%
%

\bibliographystyle{IEEEtran}
\bibliography{IEEEabrv,MIRL_database_init_ieee}

\end{document}